\documentclass[aip, jmp]{revtex4-1}

\usepackage[utf8]{inputenc}
\usepackage[T1]{fontenc}
\usepackage{mathptmx}

\usepackage{amsmath}
\usepackage{amssymb}
\usepackage{mathrsfs}
\usepackage{braket}
\usepackage{bm}

\usepackage{latexsym}
\def\qed{\\ \hfill $\Box$}
\usepackage{ntheorem}
\theoremstyle{break}
\newtheorem{thm}{Theorem}[section]
\newtheorem{lem}{Lemma}[section]
\newtheorem{cor}{Corollary}[section]
\newtheorem{prop}{Proposition}[section]
\newtheorem*{thm**}{Theorem}
\newtheorem*{lem**}{Lemma}
\newtheorem*{cor**}{Corollary}
\newtheorem*{prop**}{Proposition}

\newtheorem{lemapp}{Lemma}[section]

\newtheorem{propapp}{Proposition}[section]
\newtheorem*{propref}{Proposition \ref{prop_ortho repr}}
\newtheorem*{propref1}{Proposition \ref{prop_transitive self-dual}}
\newtheorem*{propref2}{Proposition \ref{prop_error bar and Werner}}

\theorembodyfont{\normalfont}
\newtheorem{defi}{Definition}[section]
\newtheorem*{defi**}{Definition}
\newtheorem*{pf}{Proof}

\theoremheaderfont{\itshape}
\newtheorem{eg}{Example}[section]
\newtheorem*{eg**}{Example}
\newtheorem{rmk}{Remark}[section]
\newtheorem*{rmk**}{Remark}

\usepackage{graphicx}
\makeatletter
\DeclareRobustCommand\widecheck[1]{{\mathpalette\@widecheck{#1}}}
\def\@widecheck#1#2{%
	\setbox\z@\hbox{\m@th$#1#2$}%
	\setbox\tw@\hbox{\m@th$#1%
		\widehat{%
			\vrule\@width\z@\@height\ht\z@
			\vrule\@height\z@\@width\wd\z@}$}%
	\dp\tw@-\ht\z@
	\@tempdima\ht\z@ \advance\@tempdima2\ht\tw@ \divide\@tempdima\thr@@
	\setbox\tw@\hbox{%
		\raise\@tempdima\hbox{\scalebox{1}[-1]{\lower\@tempdima\box\tw@}}}%
	{\ooalign{\box\tw@ \cr \box\z@}}}
\makeatother

\newcommand{\1}{\mbox{1}\hspace{-0.25em}\mbox{l}}
\newcommand{\ketbra}[2]{\ket{#1}\hspace{-0.25em}\bra{#2}}

\newcommand{\R}{\mathbb{R}}
\newcommand{\C}{\mathbb{C}}

\newcommand{\Tr}{\mathrm{Tr}}

\newcommand{\HH}{\mathcal{H}}
\newcommand{\LL}{\mathcal{L}}

\begin{document}  
\title{Preparation Uncertainty Implies Measurement Uncertainty in a Class of Generalized Probabilistic Theories}

\author{Ryo Takakura}
\email[]{takakura.ryo.27v@st.kyoto-u.ac.jp}
\affiliation{Department of Nuclear Engineering Kyoto University, Kyoto daigaku-katsura, Nishikyo-ku, Kyoto, 615-8540, Japan}
\author{Takayuki Miyadera}
\email[]{miyadera@nucleng.kyoto-u.ac.jp}

\date{July 21. 2020}

\begin{abstract}
In quantum theory, it is known for a pair of noncommutative observables that there is no state on which they take simultaneously definite values, and that there is no joint measurement of them.
They are called preparation uncertainty and measurement uncertainty respectively, and research has unveiled that they are not independent from but related with each other in a quantitative way. 
This study aims to reveal whether similar relations to quantum ones hold also in generalized probabilistic theories (GPTs).
In particular, a certain class of GPTs is considered which can be characterized by transitivity and self-duality and regarded as extensions of quantum theory. 
It is proved that there are close connections expressed quantitatively between two types of uncertainty on a pair observables also in those theories: if preparation uncertainty exists, then measurement uncertainty also exists, and they are described by similar inequalities.
Our results manifest that their correspondences are not specific to quantum theory but more universal ones.
\end{abstract}

\pacs{03.65.Ta, 03.65.Ud}
\maketitle

\section{Introduction}
Since it was propounded by Heisenberg \cite{Heisenberg1927}, the existence of uncertainty relations, which is not observed in classical theory, has been regarded as one of the most significant features of quantum theory. The importance of uncertainty relations lies not only in their conceptual aspects but also in practical use such as the security proof of quantum key distribution \cite{Koashi_2006}. There have been researches to capture and formulate the notion of ``uncertainty'' in several ways. One of the most outstanding works was given by Robertson \cite{PhysRev.34.163}. There was shown an uncertainty relation in terms of standard derivation which stated that the probability distributions obtained by the measurements of a pair of noncommutative observables cannot be simultaneously sharp.
While this type of uncertainty (called {\it preparation uncertainty}) has been studied also in a more direct way \cite{Uffink_PhD,PhysRevA.71.052325,PhysRevA.76.062108} or the entropic way \cite{PhysRevLett.50.631,PhysRevLett.60.1103,10.2307/25051432}, another type of uncertainty called {\it measurement uncertainty} 
is known to exist in quantum theory \cite{Busch_quantummeasurement}. It describes that when we consider measuring jointly a pair of noncommutative observables, there must exist {\it measurement error} for the joint measurement, that is, we can only conduct their {\it approximate joint measurement}. 
There have been researches on measurement uncertainty with measurement error formulated in terms of standard derivation \cite{PhysRevLett.60.2447,doi:10.1002/j.1538-7305.1965.tb01684.x,PhysRevA.67.042105} or entropy \cite{PhysRevLett.112.050401}. 
Their measurement uncertainty relations were proved mathematically by using preparation uncertainty relations. 
It implies that there may be a close connection between those two kinds of uncertainty. 
From this point of view, one of us \cite{doi:10.1063/1.3614503} proved simple inequalities which demonstrate in a more explicit way than other previous studies that preparation uncertainty indicates measurement uncertainty and the bound derived from the former also bounds the latter. The main results of \onlinecite{doi:10.1063/1.3614503} were obtained with preparation uncertainty quantified by overall widths and minimum localization error, and measurement uncertainty by error bar widths, Werner's measure, and $l_{\infty}$ distance \cite{doi:10.1063/1.2759831,PhysRevA.78.052119,10.5555/2017011.2017020,10.5555/2011593.2011606}. On the other hand, researches on {\it generalized probabilistic theories} ({\it GPTs}) \cite{Gudder_stochastic,Araki_mathematical,hardy2001quantum,PhysRevA.75.032304,PhysRevA.84.012311,BARNUM20113,1751-8121-47-32-323001}, which are the most general theories of physics, have revealed that phenomena such as no-cloning and teleportation used to be regarded as peculiar to quantum theory are indeed possible in a broader class of theories \cite{PhysRevLett.99.240501,barnum2012teleportation}. 
Concerning about uncertainty, both preparation and measurement uncertainty can be introduced naturally also in GPTs.
In \onlinecite{PhysRevA.101.052104}, their formulations were proposed and connections between them were investigated in GPTs analogical with a qubit system.
It is of interest to give further research on how two types of generalized uncertainty are related with each other.

In this paper, we study the relations between two kinds of uncertainty in GPTs. 
We focus on a class of GPTs which are {\it transitive} and {\it self-dual} including finite dimensional classical and quantum theories, and demonstrate similar results to \onlinecite{doi:10.1063/1.3614503} in the GPTs: preparation uncertainty relations indicate measurement uncertainty relations.
More precisely, it is proved in a certain class of GPTs that if a preparation uncertainty relation gives some bound, then it is also a bound on the corresponding measurement uncertainty relation with the quantifications of uncertainty in \onlinecite{doi:10.1063/1.3614503} generalized to GPTs. Our results manifest that the close connections between two kinds of uncertainty exhibited in quantum theory are more universal ones.

This paper is organized as follows. In section \ref{sec2}, we give a brief review of GPTs. There are introduced fundamental descriptions of GPTs and several mathematical assumptions imposed in order to derive our main theorems. Some examples of GPTs such as classical, quantum, and regular polygon theories are also explained. In section \ref{sec3}, we introduce measures which quantify the width of a probability distribution. These measures are used for considering whether it is possible to localize jointly two probability distributions obtained by two kinds of measurement on one certain state, that is, they are used for describing preparation uncertainty. We also introduce measures quantifying measurement error by means of which we can formulate measurement uncertainty resulting from approximate joint measurements of two incompatible measurements. After the introductions of those quantifications, we present our main theorems and their proofs. In section \ref{sec4}, we conclude this paper with several discussions.

\section{GPTs}
\label{sec2}
In this section, we give a brief review of the mathematical formulation of GPTs. Our mathematical formulation and terms are mostly in accord with \onlinecite{kimura2010physical, KIMURA2010175, EPTCS195.4,muller2012unifying}, where more detailed descriptions are given. 
Note that in the remaining of this paper we restrict ourselves to theories embedded in finite dimensional vector spaces, i.e. finite dimensional GPTs.
For descriptions of more general GPTs, we recommend \onlinecite{Lami_PhD}.

\subsection{States, effects, and measurements}
\label{sec2-1}
A physical experiment is described by three procedures: to prepare an object
system, to perform a measurement, and to obtain a probability distribution onto the outcome values of the measurement \cite{Araki_mathematical}. Each theory of GPTs gives an intuitive description of physical experiments.

\if
In each theory of GPTs, preparation procedures are called {\it states}. The set of all states is represented by a nonempty compact convex set $\Omega$, which we call the {\it state space}, in some locally convex Hausdorff topological vector space $V$ on $\R$. For simplicity, we assume in this paper that $V$ is finite dimensional. Let us denote the affine hull of $\Omega$ by $\mathit{aff}(\Omega):=\{\sum_{i=1}^{k}\theta_{i}\omega_{i}\mid k\in\mathbb{N}, \omega_{i}\in\Omega, \theta_{i}\in\R, \sum_{i=1}^{k}\theta_{i}=1\}$, and assume in the remaining of this paper that $\mathit{aff}(\Omega)$ is a $N$-dimensional ($N<\infty$) affine space, i.e. $\mathrm{dim}\mathit{aff}(\Omega)=N$\footnote{The affine dimension of an affine set $X$ is defined by the dimension of the set $X-x_{0}\ (x_{0}\in X)$ as a vector space.}.
We remark that the convex structure of $\Omega$ is derived from the notion of probability mixtures of states: if $\omega_{1}, \omega_{2}\in\Omega$, then $\omega:=p\omega_{1}+(1-p)\omega_{2}\in\Omega$ for $p\in[0, 1]$, where $\omega$ means the state obtained by the mixture of $\omega_{1}$ and $\omega_{2}$ with probability weights $\{p, 1-p\}$. Since $\Omega$ is a compact convex set, thanks to the Krein-Milman theorem \cite{Conway_functionalanalysis}, there exist extreme points of $\Omega$ which generate the whole set. We denote the set of all extreme points of $\Omega$ by $\Omega^{\mathrm{ext}}:=\{\omega_{i}^{\mathrm{ext}}\}_{i}\ (\neq\emptyset)$, and call its elements {\it pure states} (the other states are called {\it mixed states}).

In GPTs, measurements are defined through the notion of {\it effects}. Let us consider a GPT whose state space is $\Omega$, and let $\mathcal{A}(\Omega)$ be the set of all affine functions on $\Omega$, that is, $\mathcal{A}(\Omega)=\{h:\Omega\to\R\mid h(p\omega_{1}+(1-p)\omega_{2})=ph(\omega_{1})+(1-p)h(\omega_{2})\ \mbox{for all}\ p\in[0, 1],\  \omega_{1}, \omega_{2}\in\Omega \}$. An affine function $e\in\mathcal{A}(\Omega)$ is called an {\it effect} if $0\le e(\omega)\le 1$ for all $\omega\in\Omega$, where $e(\omega)$ represents the probability of obtaining a specific outcome when $\omega$ is prepared. We call the set of all effects $\mathcal{E}(\Omega)=\{e\in\mathcal{A}(\Omega)\mid0\le e(\omega)\le1\ \mbox{for all}\ \omega\in\Omega \}$ the {\it effect space} of the theory. Note that in this paper we assume the {\it no-restriction hypothesis} that all effects are allowed physically \cite{PhysRevA.81.062348}. The {\it unit effect} $u\in\mathcal{E}(\Omega)$ is defined as the effect satisfying $u(\omega)=1$ for all $\omega\in\Omega$. It can be easily shown that $\mathcal{E}(\Omega)$ is convex (the extreme elements are called {\it pure effects}), the unit effect $u$ is pure, and $u-e\in\mathcal{E}(\Omega)$ whenever $e\in\mathcal{E}(\Omega)$. An effect $e$ is called {\it indecomposable} if $e\neq0$ and a decomposition $e=e_{1}+e_{2}$, where $e_{1}, e_{2}\in\mathcal{E}(\Omega)$, implies that both $e_{1}$ and $e_{2}$ are scalar multiples of $e$. It can be seen that there exist pure and indecomposable effects in $\mathcal{E}(\Omega)$, and we denote the set of all pure and indecomposable effects by $\mathcal{E}^{\mathrm{ext}}(\Omega)=\{e_{i}^{\mathrm{ext}}\}_{i}$. In quantum theory, they correspond to rank-1 projections (see Example \ref{eg_QT} in subsection \ref{sec2-3}). A {\it measurement} or {\it observable} (with $n$ outcomes) is defined by an $n$-tuples $\{e_{i}\}_{i=1}^{n}$ of effects such that $\sum_{i=1}^{n}e_{i}=u$, where $e_{i}(\omega)$ represents the probability of observing the $i$th outcome of the measurement when a state $\omega$ is prepared. The condition $\sum_{i=1}^{n}e_{i}=u$ ensures that the total probability is 1. We describe a measurement $E$ also as $E=\{e_{a}\}_{a\in A}$ satisfying $\sum_{a\in A}^{}e_{a}=u$ in this paper, where $A$ is a sample space, namely the set of outcomes possible to be observed when $E$ is measured. In this case, $e_{a}(\omega)$ means the probability of observing the value $a\in A$ in the measurement of $E$ on a state $\omega$. We assume in this paper that all measurements are with finite outcomes and composed of nonzero effects, and do not consider the trivial measurement $\{u\}$.

It is possible to represent a GPT in another way. GPTs with state spaces $\Omega_{1}$ and $\Omega_{2}$ are called equivalent if there exists an affine bijection (affine isomorphism) $\psi$ such that $\psi(\Omega_{1})=\Omega_{2}$. It is easy to show that if  $\Omega_{1}$ and $\Omega_{2}$ are equivalent with an affine isomorphism $\psi$, then $\mathcal{E}(\Omega_{2})=\mathcal{E}(\Omega_{1})\circ\psi^{-1}$, and thus physical predictions are covariant (equivalent) in those GPTs. This allows us to assume that in a GPT with its state space $\Omega\subset V$ the affine hull of $\Omega$ does not include the origin $O$ of $V$, that is, $O\notin{\it aff}(\Omega)$. We also assume for mathematical convenience that the dimension of the embedding vector space $V$ satisfies ${\rm dim}V=N+1$ (remember that $N$ is the dimension of ${\it aff}(\Omega)$), and thus we can set $V=\R^{N+1}$ with the standard Euclidean inner product $(\cdot ,\cdot )_{E}$ because any finite dimensional Hausdorff topological vector space is isomorphic linearly and topologically to the Euclidean space with the same dimension \cite{schaefer_TopVecSp}. Note that by virtue of letting $O\notin{\it aff}(\Omega)$ and ${\rm dim}V=N+1$, any affine function on $\Omega$ can be extended uniquely to a linear function on $V$, so $\mathcal{E}(\Omega)\subset V^{\ast}\cong\R^{N+1}$, where $V^{*}$ is the dual space of $V$. 

For a state space $\Omega$, we define the set $V_{+}\subset V$ as $V_{+}:=\{x\in V\mid x=\lambda\omega,\ \omega\in\Omega, \lambda\ge0\}$ and call $V_{+}$ the {\it positive cone} generated by $\Omega$. Physically, $V_{+}$ represents the set of all ``unnormalized'' states, which are not necessarily mapped to 1 by $u$. We also define the cone $V^{*}_{+}$ dual to $V_{+}$ as $V^{*}_{+}:=\{y\in V^{*}\mid y(x)\ge0\ \mbox{for all}\ x\in V_{+}\}$. It is obvious that $\mathcal{E}(\Omega)=V^{*}_{+}\cap(u-V^{*}_{+})$, and an effect $e\in\mathcal{E}(\Omega)$ is indecomposable if and only if $e$ is on an extremal ray of $V^{*}_{+}$\footnote{A ray $E\subset V_{+}^{*}$ is called an {\it extremal ray} of $V^{*}_{+}$ if $x\in E$ and $x=y+z$ with $y, z\in V^{*}_{+}$ imply $y, z\in E$.}.
\fi

In each theory of GPTs, preparation procedures are called {\it states}. The set of all states is represented by a nonempty compact convex set $\Omega$, which we call the {\it state space}, in some locally convex Hausdorff topological vector space $V$ on $\R$. 
Let us denote the affine hull of $\Omega$ by $\mathit{aff}(\Omega):=\{\sum_{i=1}^{k}\theta_{i}\omega_{i}\mid k\in\mathbb{N}, \omega_{i}\in\Omega, \theta_{i}\in\R, \sum_{i=1}^{k}\theta_{i}=1\}$, and assume in the remaining of this paper that $\mathit{aff}(\Omega)$ is a $N$-dimensional ($N<\infty$) affine space, i.e. $\mathrm{dim}\mathit{aff}(\Omega)=N$
\footnote{The affine dimension of an affine set $X$ is defined by the dimension of the set $X-x_{0}\ (x_{0}\in X)$ as a vector space.}.
In fact, we can suppose without loss of generality that the embedding vector space $V$ satisfies ${\rm dim}V=N+1$ and ${\it aff}(\Omega)$ does not include the origin $O$ of $V$, i.e. $O\notin{\it aff}(\Omega)$\cite{Lami_PhD,Kuramochi_GPT}.
Thus, we can set $V=\R^{N+1}$ with the standard Euclidean inner product $(\cdot ,\cdot )_{E}$ because any finite dimensional Hausdorff topological vector space is isomorphic linearly and topologically to the Euclidean space with the same dimension \cite{schaefer_TopVecSp}.
We remark that the convex structure of $\Omega$ is derived from the notion of probability mixtures of states: if $\omega_{1}, \omega_{2}\in\Omega$, then $\omega:=p\omega_{1}+(1-p)\omega_{2}\in\Omega$ for $p\in[0, 1]$, where $\omega$ means the state obtained by the mixture of $\omega_{1}$ and $\omega_{2}$ with probability weights $\{p, 1-p\}$. 
Since $\Omega$ is a compact convex set, thanks to the Krein-Milman theorem \cite{Conway_functionalanalysis}, there exist extreme points of $\Omega$ which generate the whole set. 
We denote the set of all extreme points of $\Omega$ by $\Omega^{\mathrm{ext}}:=\{\omega_{i}^{\mathrm{ext}}\}_{i}\ (\neq\emptyset)$, and call its elements {\it pure states} (the other states are called {\it mixed states}).

In GPTs, measurements are defined through the notion of {\it effects}. 
Let us consider a GPT whose state space is $\Omega\subset\R^{N+1}$, and let $\mathcal{A}(\Omega)$ be the set of all affine functions on $\Omega$, that is, $\mathcal{A}(\Omega)=\{h:\Omega\to\R\mid h(p\omega_{1}+(1-p)\omega_{2})=ph(\omega_{1})+(1-p)h(\omega_{2})\ \mbox{for all}\ p\in[0, 1],\  \omega_{1}, \omega_{2}\in\Omega \}$. 
An affine function $e\in\mathcal{A}(\Omega)$ is called an {\it effect} if $0\le e(\omega)\le 1$ for all $\omega\in\Omega$, where $e(\omega)$ represents the probability of obtaining a specific outcome when $\omega$ is prepared. 
We call the set of all effects $\mathcal{E}(\Omega)=\{e\in\mathcal{A}(\Omega)\mid0\le e(\omega)\le1\ \mbox{for all}\ \omega\in\Omega \}$ the {\it effect space} of the theory. Note that in this paper we assume the {\it no-restriction hypothesis} that all effects are allowed physically \cite{PhysRevA.81.062348}. 
The {\it unit effect} $u\in\mathcal{E}(\Omega)$ is defined as the effect satisfying $u(\omega)=1$ for all $\omega\in\Omega$. It can be easily shown that $\mathcal{E}(\Omega)$ is convex (the extreme elements are called {\it pure effects}), the unit effect $u$ is pure, and $u-e\in\mathcal{E}(\Omega)$ whenever $e\in\mathcal{E}(\Omega)$. 
An effect $e$ is called {\it indecomposable} if $e\neq0$ and a decomposition $e=e_{1}+e_{2}$, where $e_{1}, e_{2}\in\mathcal{E}(\Omega)$, implies that both $e_{1}$ and $e_{2}$ are scalar multiples of $e$. It can be seen that there exist pure and indecomposable effects in $\mathcal{E}(\Omega)$, and we denote the set of all pure and indecomposable effects by $\mathcal{E}^{\mathrm{ext}}(\Omega)=\{e_{i}^{\mathrm{ext}}\}_{i}$. In quantum theory, they correspond to rank-1 projections (see Example \ref{eg_QT} in subsection \ref{sec2-3}). 
A {\it measurement} or {\it observable} (with $n$ outcomes) is defined by an $n$-tuples $\{e_{i}\}_{i=1}^{n}$ of effects such that $\sum_{i=1}^{n}e_{i}=u$, where $e_{i}(\omega)$ represents the probability of observing the $i$th outcome of the measurement when a state $\omega$ is prepared. The condition $\sum_{i=1}^{n}e_{i}=u$ ensures that the total probability is 1. 
We describe a measurement $E$ also as $E=\{e_{a}\}_{a\in A}$ satisfying $\sum_{a\in A}^{}e_{a}=u$ in this paper, where $A$ is a sample space, namely the set of outcomes possible to be observed when $E$ is measured. 
In this case, $e_{a}(\omega)$ means the probability of observing the value $a\in A$ in the measurement of $E$ on a state $\omega$. 
We assume in this paper that all measurements are with finite outcomes and composed of nonzero effects, and do not consider the trivial measurement $\{u\}$.
Note that because $O\notin{\it aff}(\Omega)$ and ${\rm dim}V=N+1$, any affine function on $\Omega$ can be extended uniquely to a linear function on $V$, so $\mathcal{A}(\Omega)= V^{\ast}\cong\R^{N+1}$, where $V^{*}$ is the dual space of $V$
\footnote{The underlying vector space $V$ of $\Omega$ is often constructed by the dual space of $\mathcal{A}(\Omega)$:  $V=\mathcal{A}(\Omega)^{*}$\cite{Lami_PhD,Kuramochi_GPT}.}.

For a state space $\Omega$, we define the set $V_{+}\subset V$ as $V_{+}:=\{x\in V\mid x=\lambda\omega,\ \omega\in\Omega, \lambda\ge0\}$ and call $V_{+}$ the {\it positive cone} generated by $\Omega$. Physically, $V_{+}$ represents the set of all ``unnormalized'' states, which are not necessarily mapped to 1 by $u$. We also define the cone $V^{*}_{+}$ dual to $V_{+}$ as $V^{*}_{+}:=\{y\in V^{*}\mid y(x)\ge0\ \mbox{for all}\ x\in V_{+}\}$. It is obvious that $\mathcal{E}(\Omega)=V^{*}_{+}\cap(u-V^{*}_{+})$, and an effect $e\in\mathcal{E}(\Omega)$ is indecomposable if and only if $e$ is on an extremal ray of $V^{*}_{+}$\footnote{A ray $E\subset V_{+}^{*}$ is called an {\it extremal ray} of $V^{*}_{+}$ if $x\in E$ and $x=y+z$ with $y, z\in V^{*}_{+}$ imply $y, z\in E$.}.

It is possible to represent a GPT in another way. GPTs with state spaces $\Omega_{1}$ and $\Omega_{2}$ are called equivalent if there exists an affine bijection (affine isomorphism) $\psi$ such that $\psi(\Omega_{1})=\Omega_{2}$. 
We again remark that the affine isomorphism $\psi$ is indeed a linear isomorphism on the underlying vector space $V=\R^{N+1}$.
It is easy to show that if  $\Omega_{1}$ and $\Omega_{2}$ are equivalent with an affine isomorphism $\psi$, then $\mathcal{E}(\Omega_{2})=\mathcal{E}(\Omega_{1})\circ\psi^{-1}$, and thus physical predictions are covariant (equivalent) in those GPTs.

In the remaining of this paper, we follow mainly those assumptions and notations described above.

\subsection{Physical equivalence of pure states}
\label{sec2-2}
It is known that in quantum theory all pure states are physically equivalent via unitary (and antiunitary) transformations \cite{Busch_quantummeasurement}. Similar notion to this physical equivalence of pure states can be introduced also in GPTs.

Let $\Omega$ be a state space. A map $T\colon\Omega\to\Omega$ is called a {\it state automorphism} on $\Omega$ if $T$ is an affine bijection. We denote the set of all state automorphisms on $\Omega$ by $GL(\Omega)$, and say that a state $\omega_{1}\in\Omega$ is {\it physically equivalent} to a state $\omega_{2}\in\Omega$ if there exists a $T\in GL(\Omega)$ such that $T\omega_{1}=\omega_{2}$. It was shown in \onlinecite{kimura2010physical} that the physical equivalence of $\omega_{1}, \omega_{2}\in\Omega$ is equal to the existence of some unit-preserving affine bijection $T'\colon\mathcal{E}(\Omega)\to\mathcal{E}(\Omega)$ satisfying $e(\omega_{1})=T'(e)(\omega_{2})$ for all $e\in\mathcal{E}(\Omega)$, which means $\omega_{1}$ and $\omega_{2}$ have the same physical contents on measurements. Because any affine map on $\Omega$ can be extended uniquely to a linear map on $V$, it holds that $GL(\Omega)=\{T\colon V\to V\mid T:\mbox{linear, bijective},\ T(\Omega)=\Omega\}$. It is clear that $GL(\Omega)$ forms a group, and we can represent the notion of physical equivalence of pure states by means of the transitive action of $GL(\Omega)$ on $\Omega^{\mathrm{ext}}$.
\begin{defi}[Transitive state space]
	\label{def_transitive state sp}
	A state space $\Omega$ is called {\it transitive} if $GL(\Omega)$ acts transitively on $\Omega^{\mathrm{ext}}$, that is, for any pair of pure states $\omega_{i}^{\mathrm{ext}}, \omega_{j}^{\mathrm{ext}}\in\Omega^{\mathrm{ext}}$ there exists an affine bijection $T_{ji}\in GL(\Omega)$ such that $\omega_{j}^{\mathrm{ext}}=T_{ji}\omega_{i}^{\mathrm{ext}}$.
\end{defi}
We remark that the equivalence of pure states does not depend on how the theory is expressed. In fact, when $\Omega$ is a transitive state space and $\Omega':=\psi(\Omega)$ is equivalent to $\Omega$ with a linear bijection $\psi$, it is easy to check that $GL(\Omega')=\psi\circ GL(\Omega)\circ\psi^{-1}$ and $\Omega'$ is also transitive.

In the remaining of this subsection, we let $\Omega$ be a transitive state space. In a transitive state space, we can introduce successfully the maximally mixed state as a unique invariant state with respect to every state automorphism.
\begin{prop}[{\rm \onlinecite{Davies_compactconvex}}]
\label{def_max mixed state}
For a transitive state space $\Omega$, there exists a unique state $\omega_{M}\in\Omega$ (which we call the {\it maximally mixed state}) such that $T\omega_{M}=\omega_{M}$ for all $T\in GL(\Omega)$. The unique maximally mixed state $\omega_{M}$ is given by 
\[
\omega_{M}=\int_{GL(\Omega)} T\omega^{\mathrm{ext}}\ d\mu(T),
\]
where $\omega^{\mathrm{ext}}$ is an arbitrary pure state and $\mu$ is the normalized two-sided invariant Haar measure on $GL(\Omega)$.
\end{prop}
Note in Proposition \ref{def_max mixed state} that the transitivity of $\Omega$ guarantees the independence of $\omega_{M}$ on the choice of $\omega^{\mathrm{ext}}$. When $\Omega^{\mathrm{ext}}$ is finite and $\Omega^{\mathrm{ext}}=\{\omega_{i}^{\mathrm{ext}}\}_{i=1}^{n}$, $\omega_{M}$ has a simpler form
\[
\omega_{M}=\frac{1}{n}\sum_{i=1}^{n} \omega_{i}^{\mathrm{ext}}.
\]
We should recall that the action of the linear bijection $\eta:=\frac{1}{\|\omega_{M}\|_{E}}\1_{V}$ on $\Omega$ does not change the theory, where $\|\omega_{M}\|_{E}=(\omega_{M}, \omega_{M})^{1/2}_{E}$ and $\1_{V}$ is the identity map on $V$. 
Since $\eta T \eta^{-1}=T$ holds for all $T\in GL(\Omega)$, $GL(\Omega)$ is invariant under the rescaling by $\eta$, i.e. $GL(\eta(\Omega))=GL(\Omega)$. 
It follows that the unique maximally mixed state of the rescaled state space $\eta(\Omega)$ is $\frac{1}{\|\omega_{M}\|_{E}}\omega_{M}$. 
In the remaining of this paper, when a transitive state space is discussed, we apply this rescaling and assume that $\|\omega_{M}\|_{E}=1$ holds. 
This assumption makes it easy to prove our main theorems in section \ref{sec3} via Proposition \ref{prop_ortho repr} introduced in the following.

The Haar measure $\mu$ on $GL(\Omega)$ makes it possible for us to construct a convenient representation of the theory. First of all, we define an product $\langle\cdot,\cdot \rangle_{GL(\Omega)}$ on $V$ as
\[
\langle x, y\rangle_{GL(\Omega)}:=\int_{GL(\Omega)}(Tx, Ty)_{E}\ d\mu(T)\quad (x, y\in V).
\]
Remark that in this paper we adopt $(\cdot,\cdot)_{E}$ as the reference inner product of $\langle\cdot,\cdot \rangle_{GL(\Omega)}$ although the following discussion still holds even if it is not $(\cdot,\cdot)_{E}$. 
Thanks to the properties of the Haar measure $\mu$, it holds that 
\[
\langle Tx, Ty\rangle_{GL(\Omega)}=\langle x, y\rangle_{GL(\Omega)} \quad \ ^{\forall} T\in GL(\Omega),
\]
which proves any $T\in GL(\Omega)$ to be an orthogonal transformation on $V$ with respect to the inner product $\langle\cdot,\cdot \rangle_{GL(\Omega)}$. Therefore, together with the transitivity of $\Omega$, we can see that all pure states of $\Omega$ are of equal norm, that is,
\begin{equation}
\label{eq_equal norm}
\begin{aligned}
\|\omega_{i}^{\mathrm{ext}}\|_{GL(\Omega)}
&=\langle\omega_{i}^{\mathrm{ext}}, \omega_{i}^{\mathrm{ext}}\rangle^{1/2}_{GL(\Omega)}\\
&=\langle T_{i0}\omega_{0}^{\mathrm{ext}}, T_{i0}\omega_{0}^{\mathrm{ext}}\rangle^{1/2}_{GL(\Omega)}\\
&=\langle\omega_{0}^{\mathrm{ext}}, \omega_{0}^{\mathrm{ext}}\rangle^{1/2}_{GL(\Omega)}\\
&=\|\omega_{0}^{\mathrm{ext}}\|_{GL(\Omega)}
\end{aligned}
\end{equation}
holds for all $\omega_{i}^{\mathrm{ext}}\in\Omega^{\mathrm{ext}}$, where $\omega_{0}^{\mathrm{ext}}$ is an arbitrary reference pure state. We remark that when $\|\omega_{M}\|_{E}=1$, we can obtain from the invariance of $\omega_{M}$ for $GL(\Omega)$
\begin{align*}
\|\omega_{M}\|_{GL(\Omega)}^{2}
&=\int_{GL(\Omega)}(T\omega_{M}, T\omega_{M})_{E}\ d\mu(T)\\
&=\int_{GL(\Omega)}(\omega_{M}, \omega_{M})_{E}\ d\mu(T)\\
&=\|\omega_{M}\|_{E}^{2}\int_{GL(\Omega)}\ d\mu(T)\\
&=\|\omega_{M}\|_{E}^{2},
\end{align*}
and thus $\|\omega_{M}\|_{GL(\Omega)}=1$ . The next proposition allows us to give a useful representation of the theory (the proof is given in Appendix \ref{appA}).
\begin{prop}
\label{prop_ortho repr}
For a transitive state space $\Omega$, there exists a basis $\{v_{l}\}_{l=1}^{N+1}$ of $V$ orthonormal with respect to the inner product  $\langle\cdot,\cdot\rangle_{GL(\Omega)}$ such that $v_{N+1}=\omega_{M}$ and 
\[
x\in\mathit{aff}(\Omega)\iff x=\sum_{l=1}^{N}a_{l}v_{l}+v_{N+1}=\sum_{l=1}^{N}a_{l}v_{l}+\omega_{M}\ (a_{1}, \cdots, a_{N}\in\R).
\]
\end{prop}
By employing the representation shown in Proposition \ref{prop_ortho repr}, an arbitrary $x\in\mathit{aff}(\Omega)$ can be written as a vector form that
\begin{equation}
\label{eq_vec repr}
x=
\left(
\begin{array}{c}
\bm{x} \\
1
\end{array}
\right)
\quad
\mbox{with}
\quad
\omega_{M}=
\left(
\begin{array}{c}
\bm{0} \\
1
\end{array}
\right),
\end{equation}
where the vector $\bm{x}$ is sometimes called the {\it Bloch vector} \cite{muller2012unifying,PhysRevLett.108.130401} corresponding to $x$.

\subsection{Self-duality}
\label{sec2-3}
In this part, we introduce the notion of self-duality, which plays an important role in our work. We also describe some examples of GPTs with relevant structures to transitivity or self-duality.

Let $V_{+}$ be the positive cone generated by a state space $\Omega$. We define the {\it internal dual cone} of $V_{+}$ relative to an inner product $(\cdot ,\cdot )$ on $V$ as $V^{*int}_{+(\cdot ,\cdot )}:=\{y\in V\mid(x, y)\ge0,\ ^{\forall} x\in V_{+}  \}$, which is isomorphic to the dual cone $V^{*}_{+}$ because of the Riesz representation theorem \cite{Conway_functionalanalysis}.  The self-duality of $V_{+}$ can be defined as follows.
\begin{defi}[Self-duality]
\label{def_self-duality}
$V_{+}$ is called {\it self-dual} if there exists an inner product $(\cdot ,\cdot )$ on $V$ such that $V_{+}=V^{*int}_{+(\cdot ,\cdot )}$.
\end{defi}
We remark similarly to Definition \ref{def_transitive state sp} that if $V_{+}$ generated by a state space $\Omega$ is self-dual, then the cone $V'_{+}$ generated by $\Omega':=\psi(\Omega)$ with a linear bijection $\psi$ (i.e. $V'_{+}=\psi(V_{+})$) is also self-dual. In fact, we can confirm that if $V_{+}=V^{*int}_{+(\cdot,\cdot)}$ holds for some inner product $(\cdot,\cdot)$, then $V'_{+}=V^{'*int}_{+(\cdot,\cdot)'}$ holds, where the inner product $(\cdot,\cdot)'$ is defined as $(x,y)'=(\psi^{-1}x,\  \psi^{-1}y)\ \ (x, y\in V)$.

Let us consider the case when $\Omega$ is transitive and $V_{+}$ is self-dual with respect to the inner product $\langle\cdot ,\cdot \rangle_{GL(\Omega)}$. Since $V_{+}=V^{*int}_{+\langle\cdot ,\cdot \rangle_{GL(\Omega)}}$, we can regard $V_{+}$ also as the set of unnormalized effects. In particular, every pure state $\omega_{i}^{\mathrm{ext}}\in\Omega^{\mathrm{ext}}$ can be considered as an unnormalized effect, and if we define
\begin{equation}
\label{def_indecomp effect0}
e_{i}:=\frac{\omega_{i}^{\mathrm{ext}}}{\|\omega_{i}^{\mathrm{ext}}\|_{GL(\Omega)}^{2}}=\frac{\omega_{i}^{\mathrm{ext}}}{\|\omega_{0}^{\mathrm{ext}}\|_{GL(\Omega)}^{2}},
\end{equation}
then from Cauchy-Schwarz inequality
\begin{align*}
\langle e_{i},\omega_{k}^{\mathrm{ext}}\rangle_{GL(\Omega)}
&\le\|e_{i}\|_{GL(\Omega)}\|\omega_{k}^{\mathrm{ext}}\|_{GL(\Omega)}=1
\end{align*}
holds for any pure state $\omega_{k}^{\mathrm{ext}}\in\Omega^{\mathrm{ext}}$ (thus $e_{i}$ is indeed an effect). 
The equality holds if and only if $\omega_{k}^{\mathrm{ext}}$ is parallel to $e_{i}$, i.e. $\omega_{k}^{\mathrm{ext}}=\omega_{i}^{\mathrm{ext}}$, and we can also conclude that an effect is pure and indecomposable if and only if it is of the form defined as \eqref{def_indecomp effect0} together with the fact that effects on the extremal rays of $V^{*int}_{+\langle\cdot ,\cdot \rangle_{GL(\Omega)}}=V_{+}$ are indecomposable (for more details see \onlinecite{KIMURA2010175}):
\begin{equation}
\label{def_indecomp effect}
e_{i}=\frac{\omega_{i}^{\mathrm{ext}}}{\|\omega_{i}^{\mathrm{ext}}\|_{GL(\Omega)}^{2}}=\frac{\omega_{i}^{\mathrm{ext}}}{\|\omega_{0}^{\mathrm{ext}}\|_{GL(\Omega)}^{2}}\equiv e_{i}^{\mathrm{ext}}\in\mathcal{E}^{\mathrm{ext}}(\Omega).
\end{equation}
When $|\Omega^{\mathrm{ext}}|<\infty$, it is sufficient for the discussion above that $\Omega$ is transitive and self-dual with respect to an arbitrary inner product.
\begin{prop}
	\label{prop_transitive self-dual}
	Let $\Omega$ be transitive with $|\Omega^{\mathrm{ext}}|<\infty$ and $V_+$ be self-dual with respect to some inner product. There exists a linear bijection $\Xi\colon V\to V$ such that $\Omega':=\Xi\Omega$ is transitive and the generating positive cone $V'_{+}$ is self-dual with respect to $\langle\cdot,\cdot\rangle_{GL(\Omega')}$, i.e.
	$V^{'}_+ = V_{+\langle\cdot ,\cdot \rangle_{GL(\Omega')}}^{'*int}$.
	\end{prop}
The proof is given in Appendix \ref{appB}. Proposition \ref{prop_transitive self-dual} reveals that if a theory with finite pure states is transitive and self-dual, then the theory can be expressed in the way it is self-dual with respect to $\langle\cdot,\cdot \rangle_{GL(\Omega)}$.

In the following, we present some examples of GPTs with transitivity or self-duality.
\begin{eg}[Finite dimensional classical theories]
\label{eg_CT}
Let us denote by $\Omega_{\mathrm{CT}}$ the state space of a finite dimensional classical system. $\Omega_{\mathrm{CT}}$ can be represented by means of some finite $N\in\mathbb{N}$ as the set of all probability distributions (probability vectors) $\{\mathbf{p}=(p_{1},\ \cdots,\ p_{N+1})\}\subset V=\R^{N+1}$ on some sample space $\{a_{1},\ \cdots,\ a_{N+1}\}$, i.e. $\Omega_{\mathrm{CT}}$ is the $N$-dimensional standard simplex. It is easy to justify that the set of all pure states $\Omega_{\mathrm{CT}}^{\mathrm{ext}}$ is given by $\Omega_{\mathrm{CT}}^{\mathrm{ext}}=\{\mathbf{p}_{i}^{\mathrm{ext}}\}_{i=1}^{N+1}$, where $\mathbf{p}_{i}^{\mathrm{ext}}$ is the probability distribution satisfying $(\mathbf{p}_{i}^{\mathrm{ext}})_{j}=\delta_{ij}$, and the positive cone $V_{+}$ by $V_{+}=\{\sigma=(\sigma_{1},\cdots,\sigma_{N+1})\in V\mid \sigma_{i}\ge0,\ ^{\forall} i\}$. Remark that the set
\[
\{\mathbf{p}_{i}^{\mathrm{ext}}\}_{i=1}^{N+1}=\{(1, 0, \cdots, 0), (0, 1, \cdots, 0), \cdots, (0, 0, \cdots, 1)\}
\]
forms a standard orthonormal basis of $V$.
Since any state automorphism maps pure states to pure states, it can be seen that the set $GL(\Omega_{\mathrm{CT}})$ of all state automorphisms on $\Omega_{\mathrm{CT}}$ is exactly the set of all permutation matrices with respect to the orthonormal basis $\{\mathbf{p}_{i}^{\mathrm{ext}}\}_{i=1}^{N+1}$ of $V$. Therefore, $\Omega_{\mathrm{CT}}$ is a transitive state space, and any $T\in GL(\Omega_{\mathrm{CT}})$ is orthogonal, which results in
\begin{align}
\label{eg_eq_CT_inner product}
\langle x, y\rangle_{GL(\Omega_{\mathrm{CT}})}
&=\int_{GL(\Omega_{\mathrm{CT}})}(Tx, Ty)_{E}\ d\mu(T)\notag\\
&=\int_{GL(\Omega_{\mathrm{CT}})}(x, y)_{E}\ d\mu(T)\notag\\
&=(x, y)_{E}\int_{GL(\Omega_{\mathrm{CT}})}d\mu(T)\notag\\
&=(x, y)_{E}.
\end{align}
The set of all positive linear functions on $\Omega_{\mathrm{CT}}$ can be identified with the internal dual cone  $V^{*int}_{+(\cdot ,\cdot )_{E}}$, and any $h\in V^{*int}_{+(\cdot ,\cdot )_{E}}$ can be represented as $h=(h(\mathbf{p}_{1}^{\mathrm{ext}}),\ \cdots,\ h(\mathbf{p}_{N+1}^{\mathrm{ext}}))$ with all entries nonnegative since
\[
h(\mathbf{p}_{i}^{\mathrm{ext}})=(h, \mathbf{p}_{i}^{\mathrm{ext}})_{E}=(h)_{i}\ge0
\]
holds for all $i$. Therefore, we can conclude together with \eqref{eg_eq_CT_inner product} $V_{+}=V^{*int}_{+(\cdot ,\cdot )_{E}}=V^{*int}_{+\langle\cdot ,\cdot \rangle_{GL(\Omega_{\mathrm{CT}})}}$. Note that we can find the representation \eqref{eq_vec repr} to be valid for this situation
by taking a proper basis of $V=\R^{N+1}$ and normalization.
\end{eg}

\begin{eg}[Finite dimensional quantum theories]
\label{eg_QT}
The state space of a finite dimensional quantum system denoted by $\Omega_{\mathrm{QT}}$ is the set of all density operators on $N(<\infty)$ dimensional Hilbert space $\HH$, that is, $\Omega_{\mathrm{QT}}:=\{\rho\in\LL_{S}(\HH)\mid\rho\ge0, \Tr[\rho]=1\}$, where $\LL_{S}(\HH)$ is the set of all self-adjoint operators on $\HH$. The set of all pure states $\Omega_{\mathrm{QT}}^{\mathrm{ext}}$ is given by the rank-1 projections: $\Omega_{\mathrm{QT}}^{\mathrm{ext}}=\{\ketbra{\psi}{\psi}\mid\ket{\psi}\in\HH, \braket{\psi|\psi}=1\}$. It has been demonstrated in \onlinecite{20639e6033144587b5fc25dd02933234} that with the identity operator $\1_{N}$ on $\HH$ and the generators $\{\sigma_{i}\}_{i=1}^{N^{2}-1}$ of $SU(N)$ satisfying
\begin{equation}
\label{eq_Pauli}
\sigma_{i}\in\LL_{S}(\HH),\ \ \Tr[\sigma_{i}]=0,\ \ \Tr[\sigma_{i}\sigma_{j}]=2\delta_{ij},
\end{equation}
any $A\in\LL_{S}(\HH)$ can be represented as
\begin{equation}
\label{eq_Bloch repr1}
A=c_{0}\1_{N}+\sum_{i=1}^{N^{2}-1}c_{i}\sigma_{i}\quad(c_{0}, c_{1}, \cdots, c_{N^{2}-1}\in\R)
\end{equation}
and any $B\in\mathit{aff}(\Omega_{\mathrm{QT}})$ as
\begin{equation}
\label{eq_Bloch repr2}
B=\frac{1}{N}\1_{N}+\sum_{i=1}^{N^{2}-1}c_{i}\sigma_{i}\quad(c_{1}, \cdots, c_{N^{2}-1}\in\R).
\end{equation}
Since \eqref{eq_Pauli} implies that $\{\1_{N}, \sigma_{1}, \cdots, \sigma_{N^{2}-1}\}$ forms an orthogonal basis of $\LL_{S}(\HH)$ with respect to the Hilbert-Schmidt inner product $(\cdot ,\cdot )_{HS}$ defined by
\[
(X,Y)_{HS}=\Tr[X^{\dagger}Y],
\]
and \eqref{eq_Bloch repr1} and \eqref{eq_Bloch repr2} prove $\mathrm{dim}(\LL_{S}(\HH))=\mathrm{dim}(\mathit{aff}(\Omega_{\mathrm{QT}}))+1$, it seems natural to consider $\Omega_{\mathrm{QT}}$ to be embedded in $V=\LL_{S}(\HH)$ equipped with $(\cdot ,\cdot )_{HS}$. Because it holds that 
\begin{align*}
\mathcal{E}(\Omega_{\mathrm{QT}})
&=\{E\in\LL_{S}(\HH)\mid0\le\Tr[E\rho]\le1,\ ^{\forall}\rho\in\Omega_{\mathrm{QT}}\}\\
&=\{E\in\LL_{S}(\HH)\mid0\le E\le\1_{N}\},
\end{align*}
we can see $V_{+}=V^{*int}_{+(\cdot ,\cdot )_{HS}}=\{A\in\LL_{S}(\HH)\mid A\ge0\}$, and rank-1 projections are pure and indecomposable effects in quantum theories.

On the other hand, it is known that in quantum theory any state automorphism is either a unitary or antiunitary transformation \cite{Busch_quantummeasurement}, and for any pair of pure states one can find a unitary operator which links them. Thus, $\Omega_{\mathrm{QT}}$ is transitive, and any state automorphism is of the form
\[
\rho\mapsto U\rho U^{\dagger}\quad\ ^{\forall}\rho\in\Omega_{\mathrm{QT}},
\]
where $U$ is unitary or antiunitary. Considering that 
\begin{align*}
(UXU^{\dagger}, UYU^{\dagger})_{HS}
&=\Tr\left[UX^{\dagger}U^{\dagger}UYU^{\dagger}
\right]\\
&=\Tr[X^{\dagger}Y]\\
&=(X,Y)_{HS}
\end{align*}
holds for any unitary or antiunitary operator $U$, we can obtain in a similar way to \eqref{eg_eq_CT_inner product}
\begin{align}
\label{eg_eq_QT_inner product}
\langle X, Y\rangle_{GL(\Omega_{\mathrm{QT}})}=(X, Y)_{HS}.
\end{align}
Therefore, we can conclude $V_{+}=V^{*int}_{+(\cdot ,\cdot )_{HS}}=V^{*int}_{+\langle \cdot ,\cdot \rangle_{GL(\Omega_{\mathrm{QT}})}}$. We remark similarly to the classical cases that we may rewrite \eqref{eq_Bloch repr2} as \eqref{eq_vec repr} by taking a suitable normalization and considering that $\omega_{M}=\1_{N}/N$.
\end{eg}

\begin{eg}[Regular polygon theories]
\label{eg_polygon}
If the state space of a GPT is in the shape of a regular polygon with $n(\ge3)$ sides, then we call it a {\it regular polygon theory} and denote the state space by $\Omega_{n}$. We set $V=\R^{3}$ when considering regular polygon theories, and it can be seen in \onlinecite{1367-2630-13-6-063024} that the pure states of $\Omega_{n}$ are described as
\[
\Omega^{\mathrm{ext}}_{n}=\{\omega_{n}^{\mathrm{ext}}(i)\}_{i=0}
^{n-1}
\]
with
\begin{align}
\label{def_polygon pure state}
\omega_{n}^{\mathrm{ext}}(i)=
\left(
\begin{array}{c}
r_{n}\cos({\frac{2\pi i}{n}})\\
r_{n}\sin({\frac{2\pi i}{n}})\\
1
\end{array}
\right),\ \ &r_{n}=\sqrt{\frac{1}{\cos({\frac{\pi}{n}})}}
\end{align}
when $n$ is finite, and when $n=\infty$ (the state space $\Omega_{\infty}$ is a disc), 
\[
\Omega_{\infty}^{\mathrm{ext}}=\{\omega_{\infty}^{\mathrm{ext}}(\theta)\}_{\theta\in[0, 2\pi)}
\]
with
\begin{align}
\label{def_disc pure state}
\omega_{\infty}^{\mathrm{ext}}(\theta)=
\left(
\begin{array}{c}
\cos\theta\\
\sin\theta\\
1
\end{array}
\right).
\end{align}
The state space $\Omega_{3}$ represents a classical trit system (the 2-dimensional standard simplex), while $\Omega_{\infty}$ represents a qubit system with real coefficients since the unit disc can be considered to be an equatorial plane of the Bloch ball. Regular polygon theories can be regarded as intermediate theories of those theories \cite{Takakura_2019}.

The state space of the regular polygon theory with $n$ sides (including $n=\infty$) defines its positive cone $V_{+}$, and it is also shown in \onlinecite{1367-2630-13-6-063024} that the corresponding internal dual cone $V^{*int}_{+(\cdot ,\cdot )_{E}}\subset\R^{3}$ is given by the conic hull\footnote{The conic hull of a set X is defined by $\mathit{cone}(X):=\{\sum_{i=1}^{k}\theta_{i}x_{i}\mid k\in\mathbb{N}, x_{i}\in X, \theta_{i}\ge0\}$.} of the following extreme effects (in fact, those effects are also indecomposable)
\begin{equation}
\label{def_polygon pure effect}
\begin{aligned}
&e_{n}^{\mathrm{ext}}(i)=\frac{1}{2}
\left(
\begin{array}{c}
r_{n}\cos({\frac{(2i-1)\pi}{n}})\\
r_{n}\sin({\frac{(2i-1)\pi}{n}})\\
1
\end{array}
\right),\ \ i=0, 1, \cdots, n-1\ \ (n:\mbox{even})\ ;\\
&e_{n}^{\mathrm{ext}}(i)=\frac{1}{1+r_{n}^{2}}
\left(
\begin{array}{c}
r_{n}\cos({\frac{2i\pi}{n}})\\
r_{n}\sin({\frac{2i\pi}{n}})\\
1
\end{array}
\right),\ \ i=0, 1, \cdots, n-1\ \ (n:\mbox{odd})\ ;\\
&e_{\infty}^{\mathrm{ext}}(\theta)=\frac{1}{2}
\left(
\begin{array}{c}
\cos\theta\\
\sin\theta\\
1
\end{array}
\right),\ \ \theta\in[0, 2\pi)\ \ \ (n=\infty).
\end{aligned}
\end{equation}
Moreover, for finite $n$, we can see that the group $GL(\Omega_{n})$ (named the {\it dihedral group}) is composed of orthogonal transformations with respect to $(\cdot ,\cdot )_{E}$ \cite{Dummit_abstractalgebra}, which also holds for $n=\infty$. Similar calculations to \eqref{eg_eq_CT_inner product} or \eqref{eg_eq_QT_inner product} demonstrate $(\cdot ,\cdot )_{E}=\langle\cdot ,\cdot \rangle_{GL(\Omega_{n})}$ for $n=3, 4, \cdots, \infty$. Therefore, from \eqref{def_polygon pure state} - \eqref{def_polygon pure effect}, we can conclude that $V_{+}$ is self-dual, i.e. $V_{+}=V^{*int}_{+(\cdot ,\cdot )_{E}}=V^{*int}_{+\langle \cdot ,\cdot \rangle_{GL(\Omega_{n})}}$, when $n$ is odd or $\infty$, while $V_{+}$ is not identical but only isomorphic to $V^{*int}_{+\langle \cdot ,\cdot \rangle_{GL(\Omega_{n})}}$ when $n$ is even (in that case, $V_{+}$ is called {\it weakly self-dual} \cite{barnum2012teleportation,1367-2630-13-6-063024}).
\end{eg}

\section{Preparation Uncertainty and Measurement Uncertainty in a Class of GPTs}
\label{sec3}
In this section, our main results on the relations between preparation uncertainty and measurement uncertainty are given in GPTs with transitivity and self-duality with respect to $\langle\cdot ,\cdot \rangle_{GL(\Omega)}$. Measures quantifying the width of a probability distribution or measurement error are also given to formulate those results. Throughout this section, we consider measurements whose sample spaces are finite metric spaces.
\subsection{Widths of probability distributions}
\label{sec3-1}
In this subsection, we give two kinds of measure to quantify how concentrated a probability distribution is. 

Let $A$ be a finite metric space equipped with a metric function $d_{A}$, and $O_{d_{A}}(a;\,w)$ be the ball defined by $O_{d_{A}}(a;\,w):=\{x\in A\mid d_{A}(x, a)\le w/2\}$. For $\epsilon\in[0,1]$ and a probability distribution $\mathbf{p}$ on $A$, we define the {\it overall width} (at confidence level $1-\epsilon$) \cite{doi:10.1063/1.3614503,doi:10.1063/1.2759831} as
\begin{equation}
\label{def_overall width original}
W_{\epsilon}(\mathbf{p}):=\inf\{w>0\mid\exists a\in A : \mathbf{p}(O_{d_{A}}(a;\,w))\ge1-\epsilon\}.
\end{equation}
We can give another formulation for the width of $\mathbf{p}$. We define the {\it minimum localization error} \cite{doi:10.1063/1.3614503} of $\mathbf{p}$ as
\begin{equation}
\label{def_localization error original}
LE(\mathbf{p}):=1-\underset{a\in A}{\max}\ p(a).
\end{equation}
Both \eqref{def_overall width original} and \eqref{def_localization error original} can be applied to probability distributions observed in physical experiments. Let us consider a GPT with $\Omega$ being its state space. For a state $\omega\in\Omega$ and a measurement $F=\{f_{a}\}_{a\in A}$ on $A$, we denote by $\omega^{F}$ the probability distribution obtained by the measurements of $F$ on $\omega$, i.e.
\[
\omega^{F}:=\{f_{a}(\omega)\}_{a\in A}.
\]
The overall width and minimum localization error for $\omega^{F}$ can be defined as
\begin{equation}
\label{def_overall width}
W_{\epsilon}(\omega^{F}):=\inf\{w>0\mid\exists a\in A : \sum_{a'\in O_{d_{A}}(a;\,w)}f_{a'}(\omega)\ge1-\epsilon\}
\end{equation}
and
\begin{equation}
\label{def_localization error}
LE(\omega^{F}):=1-\underset{a\in A}{\max}\ f_{a}(\omega)
\end{equation}
respectively. Note that as in \onlinecite{doi:10.1063/1.3614503,doi:10.1063/1.2759831}, overall widths can be defined properly even if the sample spaces of probability distributions are infinite. For example, overall widths are considered in \onlinecite{doi:10.1063/1.2759831} for probability measures on $\R$ derived from the measurement of position or momentum of a particle.

Those two measures above are used for the mathematical description of {\it preparation uncertainty relations (PURs)}. As a simple example, we consider a qubit system with Hilbert space $\HH=\C^{2}$. For two projection-valued measures (PVMs) $Z=\{\ketbra{0}{0}, \ketbra{1}{1}\}$ and $X=\{\ketbra{+}{+}, \ketbra{-}{-}\}$, where $\{\ket{0}, \ket{1}\}$ and $\{\ket{+}, \ket{-}\}=\{\frac{1}{\sqrt{2}}(\ket{0}+\ket{1}), \frac{1}{\sqrt{2}}(\ket{0}-\ket{1})\}$ are the $z$-basis and $x$-basis of $\HH$ respectively, it holds from \onlinecite{PhysRevA.71.052325,PhysRevLett.60.1103} that 
\begin{equation}
\label{eq_landau pollak}
LE(\rho^{Z})+LE(\rho^{X})\ge1-\frac{1}{\sqrt{2}}>0
\end{equation}
for any state $\rho$. The inequality \eqref{eq_landau pollak} shows that there is no state $\rho$ which makes both $LE(\rho^{Z})$ and $LE(\rho^{X})$ zero, that is, $\rho^{Z}$ and $\rho^{X}$ cannot be localized simultaneously even if the measurements are ideal ones (PVMs). PURs in terms of overall widths were also discussed in \onlinecite{doi:10.1063/1.2759831} for the position and momentum observables.

\subsection{Measurement error}
\label{sec3-2}
In this subsection, we introduce the concept of measurement error in GPTs, which derives from joint measurement problems, and describe how to quantify it. 

Let us consider a GPT with its state space $\Omega$, and two measurements $F=\{f_{a}\}_{a\in A}$ and $G=\{g_{b}\}_{b\in B}$ on $\Omega$. We call $F$ and $G$ are {\it jointly measurable} ({\it compatible}) if there exists a {\it joint measurement} $M^{FG}=\{m^{FG}_{ab}\}_{(a, b)\in A\times B}$ of $F$ and $G$ satisfying 
\begin{align*}
&\sum_{b\in B}m^{FG}_{ab}=f_{a}\ \ \mbox{for all $a\in A$}\\
&\sum_{a\in A}m^{FG}_{ab}=g_{b}\ \ \mbox{for all $b\in B$},
\end{align*}
and if $F$ and $G$ are not jointly measurable, then they are called {\it incompatible} \cite{Heinosaari_2016,Busch_2013}. It was shown in \onlinecite{PhysRevA.94.042108} that all measurements are jointly measurable if and only if the theory is a simplex, i.e. a classical theory. Thus, in most GPTs, there exist pairs of measurements which are incompatible, but we can nevertheless conduct their {\it approximate joint measurements} allowing {\it measurement error}. Assume that $F$ and $G$ are incompatible. It is known that one way to compose their approximate joint measurement is adding some trivial noise to them. To see this, we consider as a simple example the incompatible pair of measurements $Z=\{\ketbra{0}{0}, \ketbra{1}{1}\}$ and $X=\{\ketbra{+}{+}, \ketbra{-}{-}\}$ in a qubit system described in the last subsection. It was demonstrated in \onlinecite{PhysRevA.87.052125} that the measurements
\begin{equation}
\label{def_appro Z and X}
\begin{aligned}
\widetilde{Z}^{\lambda}:&=\lambda Z+(1-\lambda)I\\
&=\left\{\lambda\ketbra{0}{0}+\frac{1-\lambda}{2}\1_{2},\  \lambda\ketbra{1}{1}+\frac{1-\lambda}{2}\1_{2}\right\}\\
\widetilde{X}^{\lambda}:&=\lambda X+(1-\lambda)I\\
&=\left\{\lambda\ketbra{+}{+}+\frac{1-\lambda}{2}\1_{2},\  \lambda\ketbra{-}{-}+\frac{1-\lambda}{2}\1_{2}\right\}
\end{aligned}
\end{equation}
are jointly measurable for $0\le\lambda\le\frac{1}{\sqrt{2}}$, where $I:=\{\1_{2}/2, \1_{2}/2\}$ is a trivial measurement. The joint measurablity of \eqref{def_appro Z and X} implies that the addition of trivial noise described by a trivial observable makes incompatible measurements compatible in an approximate way. In fact, it is observed also in GPTs that adding trivial noise results in approximate joint measurements of incompatible measurements \cite{Busch_2013,PhysRevA.87.052125,PhysRevA.89.022123}. 

Because the notion of measurement error derives from the difference between ideal and approximate measurements as discussed above, we have to define ideal measurements in GPTs in order to quantify measurement error. In this paper, they are defined in an analogical way with the ones in finite dimensional quantum theories, where PVMs are considered to be ideal \cite{Busch_quantummeasurement}. If we denote a PVM by $E=\{P_{a}\}_{a}$, then each effect is of the form 
\[
P_{a}=\sum_{i_{(a)}}\ketbra{\psi_{i_{(a)}}}{\psi_{i_{(a)}}}.
\]
In particular, every effect is a sum of pure and indecomposable effects, and we call in a similar way a measurement $F=\{f_{a}\}_{a\in A}$ on $\Omega$ ideal if each effect $f_{a}$ satisfies
\begin{equation}
\label{def of ideal meas}
f_{a}=\sum_{i_{(a)}}e_{i_{(a)}}^{\mathrm{ext}},\quad\mbox{or}\quad f_{a}=u-\sum_{i_{(a)}}e_{i_{(a)}}^{\mathrm{ext}},
\end{equation}
where we should recall that the set of all pure and indecomposable effects is denoted by $\{e_{i}^{\mathrm{ext}}\}_{i}$ and we do not consider the trivial measurement $F=\{u\}$. It is easy to see that measurements defined as \eqref{def of ideal meas} result in PVMs in finite dimensional quantum theories. This type of measurement was considered also in \onlinecite{Barnum_2014}.

The introduction of ideal measurements makes it possible for us to quantify measurement error. Consider an ideal measurement $F=\{f_{a}\}_{a}$ and a general measurement $\widetilde{F}=\{\widetilde{f}_{a}\}_{a}$, and suppose similarly to the previous subsection that $A$ is a finite metric space with a metric $d_{A}$. 
$F$ may be understood as the measurement intended to be measured, while $\widetilde{F}$ as a measurement conducted actually.
Taking into consideration the fact that for each nonzero pure effect there exists at least one state which is mapped to 1 (an ``eigenstate'' \cite{KIMURA2010175}), we can define for $\epsilon\in[0,1]$ the {\it error bar width} of $\widetilde{F}$ relative to $F$ \cite{doi:10.1063/1.3614503,doi:10.1063/1.2759831} as
\begin{equation}
\label{def_error bar width}
\begin{aligned}
\mathcal{W}_{\epsilon}(\widetilde{F}, F)
&=\inf\{w>0\mid\ ^{\forall} a\in A, \ ^{\forall} \omega\in\Omega :\\ 
&\qquad\qquad\quad\quad
f_{a}(\omega)=1\Rightarrow\sum_{a'\in O_{d_{A}}(a;\,w)}\widetilde{f}_{a'}(\omega)\ge1-\epsilon\}.
\end{aligned}
\end{equation}
$\mathcal{W}_{\epsilon}(\widetilde{F}, F)$ represents the spread of probabilities around the ``eigenvalues'' of $F$ observed when the corresponding ``eigenstates'' of $F$ are measured by $\widetilde{F}$, and thus it can be thought to be one of the quantifications of measurement error. Note that although error bar widths in general (not necessarily finite) metric spaces were defined in \onlinecite{doi:10.1063/1.2759831}, we consider only finite metric spaces in this paper, so we employ their convenient forms \eqref{def_error bar width} in finite metric spaces shown in \onlinecite{doi:10.1063/1.3614503}. Another measure is the one given by Werner \cite{10.5555/2011593.2011606} as the difference of expectation values of ``slowly varying functions'' on the probability distributions obtained when $F$ and $\widetilde{F}$ are measured. It is defined as
\begin{equation}
\label{def_Werner's measure}
D_{W}(\widetilde{F}, F):=\underset{\omega\in\Omega}{\sup}\ \underset{h\in\Lambda}{\sup}
\left|(\tilde{F}[h])(\omega)-(F[h])(\omega)\right|,
\end{equation}
where
\[
\Lambda:=\{h\colon A\to\R\mid |h(a_{1})-h(a_{2})|\le d_{A}(a_{1}, a_{2}),\ ^{\forall} a_{1}, a_{2}\in A\}
\]
is the set of all ``slowly varying functions'' (called the {\it Lipshitz ball} of $(A, d_{A})$) and
\[
F[h]:=\sum_{a\in A}h(a)f_{a}
\]
is a map which gives the expectation value of $h\in\Lambda$ when $F$ is measured on a state $\omega$ (similarly for $\widetilde{F}[h]$). 
There is known a simple relation between \eqref{def_error bar width} and \eqref{def_Werner's measure} (the proof is given in Appendix \ref{appC}).
\begin{prop}[{\rm \onlinecite{doi:10.1063/1.3614503,doi:10.1063/1.2759831}}]
\label{prop_error bar and Werner}
Let $(A,d_{A})$ be a finite metric space, and $F=\{f_{a}\}_{a\in A}$ and $\widetilde{F}=\{\widetilde{f}_{a}\}_{a\in A}$ be an ideal and general measurement respectively. 
Then,
\[
\mathcal{W}_{\epsilon}(\widetilde{F}, F)\le\frac{2}{\epsilon}D_{W}(\tilde{F}, F)
\]
holds for $\epsilon\in(0, 1].$
\end{prop}
\if
	For $\epsilon\in(0, 1],$
	\[
	\mathcal{W}_{\epsilon}(\widetilde{F}, F)\le\frac{2}{\epsilon}D_{W}(\tilde{F}, F)
	\]
	holds for any pair of an ideal measurement $F$ and a general measurement $\widetilde{F}$.
\fi
On the other hand, there can be introduced a more intuitive quantification of measurement error called {\it $l_{\infty}$ distance} \cite{10.5555/2017011.2017020}:
\begin{equation}
\label{def_l distance}
D_{\infty}(\widetilde{F}, F):=\underset{\omega\in\Omega}{\sup}\ \underset{a\in A}{\max}\left|\widetilde{f}_{a}(\omega)-f_{a}(\omega)\right|.
\end{equation}

By means of those quantifications of measurement error above, we can formulate {\it measurement uncertainty relations (MURs)}. As an illustration, we again consider the joint measurement problem of incompatible measurements $Z$ and $X$ in a qubit system. Suppose that $\widetilde{M}^{ZX}$ is an approximate joint measurement of $Z$ and $X$, and $\widetilde{M}^{Z}$ and $\widetilde{M}^{X}$ are its marginal measurements corresponding to $Z$ and $X$ respectively. It was proved in \onlinecite{10.5555/2017011.2017020} that 
\begin{equation}
\label{eq_MUR_ZX}
D_{\infty}(\widetilde{M}^{Z}, Z)+D_{\infty}(\widetilde{M}^{X}, X)\ge 1-\frac{1}{\sqrt{2}}>0.
\end{equation}
\eqref{eq_MUR_ZX} gives a quantitative representation of the incompatibility of $Z$ and $X$ that $D_{\infty}(\widetilde{M}^{Z}, Z)$ and $D_{\infty}(\widetilde{M}^{X}, X)$ cannot be simultaneously zero, that is, measurement error must occur when conducting any approximate joint measurement of $Z$ and $X$ (see \onlinecite{PhysRevA.78.052119} for another inequality). MURs for the position and momentum observables were given in \onlinecite{doi:10.1063/1.2759831} and \onlinecite{10.5555/2011593.2011606} in terms of \eqref{def_error bar width} and \eqref{def_Werner's measure} respectively.

\subsection{Main results: relations between preparation uncertainty and measurement uncertainty}
In the previous subsections, we have introduced several measures to review two kinds of uncertainty, preparation uncertainty and measurement uncertainty. In this part, we shall manifest as our main results how they are related with each other in GPTs, which is a generalization of the quantum ones in \onlinecite{doi:10.1063/1.3614503}.

Before demonstrating our main theorems, we confirm the physical settings and mathematical assumptions to state them. In the following, we focus on a GPT with $\Omega$ being its state space, and suppose that $\Omega$ is transitive and its positive cone $V_{+}$ is self-dual with respect to $\langle\cdot ,\cdot \rangle_{GL(\Omega)}$. In addition, we consider ideal measurements $F=\{f_{a}\}_{a\in A}$ and $G=\{g_{b}\}_{b\in B}$ on $\Omega$, whose sample spaces are finite metric spaces $(A, d_{A})$ and $(B, d_{B})$ respectively, and consider a measurement $\widetilde{M}^{FG}:=\{\widetilde{m}_{ab}^{FG}\}_{(a, b)\in A\times B}
$ as an approximate joint measurement of $F$ and $G$, whose marginal measurements are given by
\[
\begin{aligned}
&\widetilde{M}^{F}:=\{\widetilde{m}_{a}^{F}\}_{a},\quad \widetilde{m}_{a}^{F}:=\sum_{b\in B}\widetilde{m}_{ab}^{FG};\\
&\widetilde{M}^{G}:=\{\widetilde{m}_{b}^{G}\}_{b},\quad \widetilde{m}_{b}^{G}:=\sum_{a\in A}\widetilde{m}_{ab}^{FG}.
\end{aligned}
\]
Remember that as shown in Subsection \ref{sec3-2} the ideal measurement $F=\{f_{a}\}_{a}$ satisfies
\begin{equation}
\label{eq_general PVM}
f_{a}=\sum_{i_{(a)}}e_{i_{(a)}}^{\mathrm{ext}},\quad\mbox{or}\quad f_{a}=u-\sum_{i_{(a)}}e_{i_{(a)}}^{\mathrm{ext}}
\end{equation}
in terms of the pure and indecomposable effects $\{e_{i}^{\mathrm{ext}}\}_{i}$ shown in \eqref{def_indecomp effect} (similarly for $G=\{g_{b}\}_{b}$).
The following lemmas are needed to prove our main results.
\begin{lem}
\label{lem_u=omegaM}
If $\Omega$ is transitive, then the unit effect $u\in V^{*int}_{+\langle\cdot ,\cdot \rangle_{GL(\Omega)}}\subset V$ is identical to the maximally mixed state $\omega_{M}$, i.e. $u=\omega_{M}$.
\end{lem}
\begin{pf}
It is an easy consequence of Proposition \ref{prop_ortho repr}. 
In fact, \eqref{eq_vec repr} gives
\[
u=\omega_{M}=
\left(
\begin{array}{c}
\bm{0} \\
1
\end{array}
\right).
\]\qed
\end{pf}

\begin{lem}
\label{lem_eigenstates}
If $\Omega$ is a transitive state space and its positive cone $V_{+}$ is self-dual with respect to $\langle\cdot,\cdot\rangle_{GL(\Omega)}$, 
then for any effect $e\in \mathcal{E}(\Omega)$ on $\Omega$ it holds that 
\begin{equation}
\label{eq_eigenstate0}
\frac{e}{\langle u, e\rangle}\in\Omega,
\end{equation}
and for any ideal measurement $F=\{f_{a}\}_{a\in A}$ on $\Omega$ it holds that
\begin{equation}
\label{eq_eigenstate}
\left\langle f_{a},\ \frac{f_{a}}{\langle u, f_{a}\rangle}\right\rangle=1
\end{equation}
for all $a\in A$. In particular, each $f_{a}/\langle u, f_{a}\rangle$ is an ``eigenstate'' of $F$.
\end{lem}
\begin{pf}
In this proof, we denote the inner product $\langle\cdot ,\cdot \rangle_{GL(\Omega)}$ and the norm $\|\cdot\|_{GL(\Omega)}$ simply by $\langle\cdot ,\cdot \rangle$ and $\|\cdot\|$ respectively.

For any element $e\in V^{*\mathrm{int}}_{+\langle\cdot ,\cdot \rangle}$, $e/\langle u, e\rangle$ defines a state
because $\left\langle u,\ e/\langle u, e\rangle\right\rangle=1$ and $e\in V_{+}$ due to the the self-duality: $V_{+}=V^{*\mathrm{int}}_{+\langle\cdot ,\cdot \rangle}$, which proves \eqref{eq_eigenstate0}.
To prove \eqref{eq_eigenstate}, we focus on the fact that $f_{a}$ in \eqref{eq_general PVM} is an effect (thus $u-f_{a}$ is also an effect), that is, $\sum_{i_{(a)}}e_{i_{(a)}}^{\mathrm{ext}}$ is an effect and it satisfies $0\le\langle\sum_{i_{(a)}}e_{i_{(a)}}^{\mathrm{ext}},\ \omega\rangle\le1$ for any state $\omega\in\Omega$. However, if we act $\sum_{i_{(a)}}e_{i_{(a)}}^{\mathrm{ext}}$ on the pure state $\omega_{j_{(a)}}^{\mathrm{ext}}$, then \eqref{def_indecomp effect} shows that $\langle e_{j_{(a)}}^{\mathrm{ext}},\ \omega_{j_{(a)}}^{\mathrm{ext}}\rangle=1$, and thus we have
\[
\langle e_{i_{(a)}}^{\mathrm{ext}},\ \omega_{j_{(a)}}^{\mathrm{ext}}\rangle=0\quad\mbox{for $i_{(a)}\neq j_{(a)}$},
\]
that is,
\begin{equation}
\label{eq_prf lemma}
\langle e_{i_{(a)}}^{\mathrm{ext}},\ e_{j_{(a)}}^{\mathrm{ext}}\rangle=0\quad\mbox{for $i_{(a)}\neq j_{(a)}$}.
\end{equation}
Because
\begin{align*}
\langle e_{i_{(a)}}^{\mathrm{ext}},\ e_{i_{(a)}}^{\mathrm{ext}}\rangle=\frac{1}{\|\omega_{0}^{\mathrm{ext}}\|^{2}}\quad\mbox{and}\quad
\langle u,\ e_{i_{(a)}}^{\mathrm{ext}}\rangle=\frac{1}{\|\omega_{0}^{\mathrm{ext}}\|^{2}}
\end{align*}
hold from \eqref{def_indecomp effect}, we obtain together with \eqref{eq_prf lemma}
\begin{equation}
\label{eq_values of indecomp effects}
\begin{aligned}
\langle\sum_{i_{(a)}}e_{i_{(a)}}^{\mathrm{ext}},\ \sum_{i_{(a)}}e_{i_{(a)}}^{\mathrm{ext}}\rangle=\frac{(\#i_{(a)})}{\|\omega_{0}^{\mathrm{ext}}\|^{2}},&\quad
\langle u,\ \sum_{i_{(a)}}e_{i_{(a)}}^{\mathrm{ext}}\rangle=\frac{(\#i_{(a)})}{\|\omega_{0}^{\mathrm{ext}}\|^{2}},\\
\langle u-\sum_{i_{(a)}}e_{i_{(a)}}^{\mathrm{ext}},\ u-\sum_{i_{(a)}}e_{i_{(a)}}^{\mathrm{ext}}\rangle=1-\frac{(\#i_{(a)})}{\|\omega_{0}^{\mathrm{ext}}\|^{2}},&\quad
\langle u,\ u-\sum_{i_{(a)}}e_{i_{(a)}}^{\mathrm{ext}}\rangle=1-\frac{(\#i_{(a)})}{\|\omega_{0}^{\mathrm{ext}}\|^{2}},
\end{aligned}
\end{equation}
where $(\#i_{(a)})$ is the number of elements of the index set $\{i_{(a)}\}$ and we use $\langle u, u\rangle=\langle u, \omega_{M}\rangle=1$ (Lemma \ref{lem_u=omegaM}). Therefore, we can conclude that every effect $f_{a}=\sum_{i_{(a)}}e_{i_{(a)}}^{\mathrm{ext}}\ \mbox{or}\  u-\sum_{i_{(a)}}e_{i_{(a)}}^{\mathrm{ext}}$ composing $F$ satisfies
\begin{equation*}
\left\langle f_{a},\ \frac{f_{a}}{\langle u, f_{a}\rangle}\right\rangle=1.
\end{equation*}\qed
\end{pf}

Now, we can state our main theorems connecting PURs and MURs. Note that one of us \cite{doi:10.1063/1.3614503} proved similar results to ours for finite dimensional quantum theories. Because GPTs shown above include those theories, our theorems can be considered to demonstrate that the relations between PURs and MURs introduced in \onlinecite{doi:10.1063/1.3614503} are more general ones.
\begin{thm}
\label{thm_error bar}
Let $\Omega$ be a transitive state space and its positive cone $V_{+}$ be self-dual with respect to $\langle\cdot,\cdot\rangle_{GL(\Omega)}$, and let $(F, G)$ be a pair of ideal measurements on $\Omega$.
For an arbitrary approximate joint measurement $\widetilde{M}^{FG}$ of $(F, G)$ and $\epsilon_{1}, \epsilon_{2}\in[0,1]$ satisfying $\epsilon_{1}+\epsilon_{2}\le1$, there exists a state $\omega\in\Omega$ such that
\[
\begin{aligned}
&\mathcal{W}_{\epsilon_{1}}(\widetilde{M}^{F}, F)\ge W_{\epsilon_{1}+\epsilon_{2}}(\omega^{F})\\
&\mathcal{W}_{\epsilon_{2}}(\widetilde{M}^{G}, G)\ge W_{\epsilon_{1}+\epsilon_{2}}(\omega^{G}).
\end{aligned}
\]
\end{thm}
Theorem \ref{thm_error bar} manifests that if one cannot make both $W_{\epsilon_{1}+\epsilon_{2}}(\omega^{F})$ and $W_{\epsilon_{1}+\epsilon_{2}}(\omega^{G})$ vanish, then one also cannot make both $\mathcal{W}_{\epsilon_{1}}(\widetilde{M}^{F}, F)$ and $\mathcal{W}_{\epsilon_{2}}(\widetilde{M}^{G}, G)$ vanish. That is, if there exists a PUR, then there also exists a MUR. Moreover, Theorem \ref{thm_error bar} also demonstrates that bounds for MURs in terms of error bar widths can be given by ones for PURs described by overall widths.

\begin{pf}[Proof of Theorem \ref{thm_error bar}]
In this proof, we denote again the inner product $\langle\cdot ,\cdot \rangle_{GL(\Omega)}$ and the norm $\|\cdot\|_{GL(\Omega)}$ simply by $\langle\cdot ,\cdot \rangle$ and $\|\cdot\|$ respectively.

\if
Since we assume that $f_{a}$ in \eqref{eq_general PVM} is an effect (thus $u-f_{a}$ is also an effect),  $\sum_{i_{(a)}}e_{i_{(a)}}^{\mathrm{ext}}$ is an effect and it satisfies $0\le\langle\sum_{i_{(a)}}e_{i_{(a)}}^{\mathrm{ext}},\ \omega\rangle\le1$ for any state $\omega\in\Omega$. However, if we act $\sum_{i_{(a)}}e_{i_{(a)}}^{\mathrm{ext}}$ on the pure state $\omega_{j_{(a)}}^{\mathrm{ext}}$, then \eqref{def_indecomp effect} shows that $\langle e_{j_{(a)}}^{\mathrm{ext}},\ \omega_{j_{(a)}}^{\mathrm{ext}}\rangle=1$, and thus we have
\[
\langle e_{i_{(a)}}^{\mathrm{ext}},\ \omega_{j_{(a)}}^{\mathrm{ext}}\rangle=0\quad\mbox{for $i_{(a)}\neq j_{(a)}$},
\]
that is,
\begin{equation}
\langle e_{i_{(a)}}^{\mathrm{ext}},\ e_{j_{(a)}}^{\mathrm{ext}}\rangle=0\quad\mbox{for $i_{(a)}\neq j_{(a)}$}.
\end{equation}
Because
\begin{align*}
\langle e_{i_{(a)}}^{\mathrm{ext}},\ e_{i_{(a)}}^{\mathrm{ext}}\rangle=\frac{1}{\|\omega_{0}^{\mathrm{ext}}\|^{2}}\quad\mbox{and}\quad
	\langle u,\ e_{i_{(a)}}^{\mathrm{ext}}\rangle=\frac{1}{\|\omega_{0}^{\mathrm{ext}}\|^{2}}
\end{align*}
hold from \eqref{def_indecomp effect}, we obtain
\begin{equation}
\begin{aligned}
\langle\sum_{i_{(a)}}e_{i_{(a)}}^{\mathrm{ext}},\ \sum_{i_{(a)}}e_{i_{(a)}}^{\mathrm{ext}}\rangle&=\frac{(\#i_{(a)})}{\|\omega_{0}^{\mathrm{ext}}\|^{2}}\\
\langle u,\ \sum_{i_{(a)}}e_{i_{(a)}}^{\mathrm{ext}}\rangle&=\frac{(\#i_{(a)})}{\|\omega_{0}^{\mathrm{ext}}\|^{2}}\\
\langle u-\sum_{i_{(a)}}e_{i_{(a)}}^{\mathrm{ext}},\ u-\sum_{i_{(a)}}e_{i_{(a)}}^{\mathrm{ext}}\rangle&=1-\frac{(\#i_{(a)})}{\|\omega_{0}^{\mathrm{ext}}\|^{2}}\\
\langle u,\ u-\sum_{i_{(a)}}e_{i_{(a)}}^{\mathrm{ext}}\rangle&=1-\frac{(\#i_{(a)})}{\|\omega_{0}^{\mathrm{ext}}\|^{2}},
\end{aligned}
\end{equation}
where $(\#i_{(a)})$ is the number of elements of the index set $\{i_{(a)}\}$ and we use $\langle u, u\rangle=\langle u, \omega_{M}\rangle=1$.

On the other hand, for any element $e\in V^{*\mathrm{int}}_{+\langle\cdot ,\cdot \rangle}$, $e/\langle u, e\rangle$ defines a state because $V_{+}=V^{*\mathrm{int}}_{+\langle\cdot ,\cdot \rangle}$. In particular, for the effect $f_{a}=\sum_{i_{(a)}}e_{i_{(a)}}^{\mathrm{ext}}\ \mbox{or}\  u-\sum_{i_{(a)}}e_{i_{(a)}}^{\mathrm{ext}}$, we can see from \eqref{eq_values of indecomp effects} that
\begin{equation}
\left\langle f_{a},\ \frac{f_{a}}{\langle u, f_{a}\rangle}\right\rangle=1
\end{equation}
holds, that is, the state $f_{a}/\langle u, f_{a}\rangle$ is an eigenstate of $f_{a}$ (similarly for $g_{b}$). 
\fi

From Lemma \ref{lem_eigenstates} and the definition of $\mathcal{W}_{\epsilon_{1}}(\widetilde{M}^{F}, F)$ \eqref{def_error bar width}, for any $w_{1}\ge\mathcal{W}_{\epsilon_{1}}(\widetilde{M}^{F}, F)$ we have
\[
\sum_{a'\in O_{d_{A}}(a;\,w_{1})}\left\langle \widetilde{m}_{a'}^{F},\ \frac{f_{a}}{\langle u, f_{a}\rangle}\right\rangle\ge1-\epsilon_{1},
\]
equivalently,
\[
\sum_{b'\in B}\sum_{a'\in O_{d_{A}}(a;\,w_{1})}\left\langle \widetilde{m}_{a'b'}^{FG},\ \frac{f_{a}}{\langle u, f_{a}\rangle}\right\rangle\ge1-\epsilon_{1}
\]
for all $a\in A$. 
Multiplying both sides by $\langle u, f_{a}\rangle=\langle \omega_{M}, f_{a}\rangle (>0)$ (Lemma \ref{lem_u=omegaM}) and taking the summation over $a$ yield
\begin{equation}
\label{eq_1}
\sum_{a\in A}\sum_{b'\in B}\sum_{a'\in O_{d_{A}}(a;\,w_{1})}\left\langle \widetilde{m}_{a'b'}^{FG},\ f_{a}\right\rangle\ge1-\epsilon_{1},
\end{equation}
where we use the relation $\sum_{a\in A}\langle u, f_{a}\rangle=\langle u, u\rangle=\langle u, \omega_{M}\rangle=1$.
Defining a function $\chi_{[d_{A}, w_{1}]}$ on $A\times A$ such that
\[
\chi_{[d_{A}, w_{1}]}(a, a')=
\left\{
\begin{aligned}
&1\qquad(d_{A}(a, a')\le\frac{w_{1}}{2})\\
&0\qquad(d_{A}(a, a')>\frac{w_{1}}{2}),
\end{aligned}
\right.
\]
it holds that 
\begin{align*}
\sum_{a\in A}\sum_{a'\in O_{d_{A}}(a;\,w_{1})}\left\langle \widetilde{m}_{a'b'}^{FG},\ f_{a}\right\rangle
&=\sum_{(a, a')\in A\times A}\chi_{[d_{A}, w_{1}]}(a, a')\left\langle \widetilde{m}_{a'b'}^{FG},\ f_{a}\right\rangle\\
&=\sum_{a'\in A}\sum_{a\in O_{d_{A}}(a';\,w_{1})}\left\langle \widetilde{m}_{a'b'}^{FG},\ f_{a}\right\rangle
\end{align*}
because of the symmetric action of $\chi_{[d_{A}, w_{1}]}$ on $a$ and $a'$.
Therefore, \eqref{eq_1} can be rewritten as
\[
\sum_{a'\in A}\sum_{b'\in B}\sum_{a\in O_{d_{A}}(a';\,w_{1})}\left\langle \widetilde{m}_{a'b'}^{FG},\ f_{a}\right\rangle\ge1-\epsilon_{1}.
\]
Overall, we obtain
\begin{equation}
\label{ineq_fa}
\sum_{a'\in A}\sum_{b'\in B}\sum_{a\in O_{d_{A}}(a';\,w_{1})}\langle u, \widetilde{m}_{a'b'}^{FG}\rangle\left\langle f_{a},\ \frac{\widetilde{m}_{a'b'}^{FG}}{\langle u, \widetilde{m}_{a'b'}^{FG}\rangle}\right\rangle\ge1-\epsilon_{1}.
\end{equation}
Similar calculations show that for any $w_{2}\ge\mathcal{W}_{\epsilon_{2}}(\widetilde{M}^{G}, G)$
\begin{equation}
\label{ineq_gb}
\sum_{a'\in A}\sum_{b'\in B}\sum_{b\in O_{d_{B}}(b';\,w_{2})}\langle u, \widetilde{m}_{a'b'}^{FG}\rangle\left\langle g_{b},\ \frac{\widetilde{m}_{a'b'}^{FG}}{\langle u, \widetilde{m}_{a'b'}^{FG}\rangle}\right\rangle\ge1-\epsilon_{2}
\end{equation}
holds.
We obtain from \eqref{ineq_fa} and \eqref{ineq_gb}
\begin{equation*}
\begin{aligned}
\sum_{a'\in A}\sum_{b'\in B}\langle u, \widetilde{m}_{a'b'}^{FG}\rangle
\left[
\left(
\sum_{a\in O_{d_{A}}(a';\,w_{1})}\left\langle f_{a},\ \frac{\widetilde{m}_{a'b'}^{FG}}{\langle u, \widetilde{m}_{a'b'}^{FG}\rangle}\right\rangle
\right)\right.\qquad\qquad\qquad\qquad\quad\\
+
\left.\left(
\sum_{b\in O_{d_{B}}(b';\,w_{2})}\left\langle g_{b},\ \frac{\widetilde{m}_{a'b'}^{FG}}{\langle u, \widetilde{m}_{a'b'}^{FG}\rangle}\right\rangle
\right)
\right]\ge 2-\epsilon_{1}-\epsilon_{2},
\end{aligned}
\end{equation*}
which implies that there exists a $(a'_{0}, b'_{0})\in A\times B$ such that 
\begin{equation}
\label{ineq_fa gb}
\begin{aligned}
\left(
\sum_{a\in O_{d_{A}}(a'_{0};\,w_{1})}\left\langle f_{a},\ \frac{\widetilde{m}_{a'_{0}b'_{0}}^{FG}}{\langle u, \widetilde{m}_{a'_{0}b'_{0}}^{FG}\rangle}\right\rangle
\right)\qquad\qquad\qquad\qquad\qquad\qquad\qquad\\
+
\left(
\sum_{b\in O_{d_{B}}(b'_{0};\,w_{2})}\left\langle g_{b},\ \frac{\widetilde{m}_{a'_{0}b'_{0}}^{FG}}{\langle u, \widetilde{m}_{a'_{0}b'_{0}}^{FG}\rangle}\right\rangle
\right)\ge 2-\epsilon_{1}-\epsilon_{2}
\end{aligned}
\end{equation}
since $\sum_{a'\in A}\sum_{b'\in B}\langle u, \widetilde{m}_{a'b'}^{FG}\rangle=\langle u, u\rangle=1$ and $0\le\langle u, \widetilde{m}_{a'b'}^{FG}\rangle\le1$ for all $(a', b')\in A\times B$. 
We can see from \eqref{ineq_fa gb} that
\begin{align}
\sum_{a\in O_{d_{A}}(a'_{0};\,w_{1})}\left\langle f_{a},\ \frac{\widetilde{m}_{a'_{0}b'_{0}}^{FG}}{\langle u, \widetilde{m}_{a'_{0}b'_{0}}^{FG}\rangle}\right\rangle
&\ge
1-\epsilon_{1}-\epsilon_{2}+\left(1-\sum_{b\in O_{d_{B}}(b'_{0};\,w_{2})}\left\langle g_{b},\ \frac{\widetilde{m}_{a'_{0}b'_{0}}^{FG}}{\langle u, \widetilde{m}_{a'_{0}b'_{0}}^{FG}\rangle}\right\rangle\right)\notag\\
\label{ineq fa epsilon}
&\ge 1-\epsilon_{1}-\epsilon_{2}
\end{align}
holds for an arbitrary $w_{1}\ge\mathcal{W}_{\epsilon_{1}}(\widetilde{M}^{F}, F)$, where we use 
\[
\sum_{b\in O_{d_{B}}(b'_{0};\,w_{2})}\left\langle g_{b},\ \frac{\widetilde{m}_{a'_{0}b'_{0}}^{FG}}{\langle u, \widetilde{m}_{a'_{0}b'_{0}}^{FG}\rangle}\right\rangle
\le
\sum_{b\in B}\left\langle g_{b},\ \frac{\widetilde{m}_{a'_{0}b'_{0}}^{FG}}{\langle u, \widetilde{m}_{a'_{0}b'_{0}}^{FG}\rangle}\right\rangle=1,
\]
and similarly
\begin{align}
\label{ineq gb epsilon}
\sum_{b\in O_{d_{B}}(b'_{0};\,w_{2})}\left\langle g_{b},\ \frac{\widetilde{m}_{a'_{0}b'_{0}}^{FG}}{\langle u, \widetilde{m}_{a'_{0}b'_{0}}^{FG}\rangle}\right\rangle
\ge
1-\epsilon_{1}-\epsilon_{2}
\end{align}
holds for an arbitrary $w_{2}\ge\mathcal{W}_{\epsilon_{2}}(\widetilde{M}^{G}, G)$. 
Because 
\[
\omega'_{0}:=\frac{\widetilde{m}_{a'_{0}b'_{0}}^{FG}}{\langle u, \widetilde{m}_{a'_{0}b'_{0}}^{FG}\rangle}
\]
defines a state (\eqref{eq_eigenstate0} in Lemma \ref{lem_eigenstates}), \eqref{ineq fa epsilon} and \eqref{ineq gb epsilon} together with the definition of the overall width \eqref{def_overall width} result in 
\begin{align*}
&w_{1}\ge W_{\epsilon_{1}+\epsilon_{2}}({\omega'}_{0}^{F})\\
&w_{2}\ge W_{\epsilon_{1}+\epsilon_{2}}({\omega'}_{0}^{G}).
\end{align*}
These equations hold for any $w_{1}\ge\mathcal{W}_{\epsilon_{1}}(\widetilde{M}^{F}, F)$ and $w_{2}\ge\mathcal{W}_{\epsilon_{2}}(\widetilde{M}^{G}, G)$, so we finally obtain
\begin{align*}
&\mathcal{W}_{\epsilon_{1}}(\widetilde{M}^{F}, F)\ge W_{\epsilon_{1}+\epsilon_{2}}({\omega'}_{0}^{F})\\
&\mathcal{W}_{\epsilon_{2}}(\widetilde{M}^{G}, G)\ge W_{\epsilon_{1}+\epsilon_{2}}({\omega'}_{0}^{G}).
\end{align*}
\qed
\end{pf}
The next corollary results immediately from Proposition \ref{prop_error bar and Werner}. It describes a similar content to Theorem \ref{thm_error bar} in terms of another measure. 
\begin{cor}
\label{cor_Werner measure}
Let $\Omega$ be a transitive state space and its positive cone $V_{+}$ be self-dual with respect to $\langle\cdot,\cdot\rangle_{GL(\Omega)}$, and let $(F, G)$ be a pair of ideal measurements on $\Omega$.
For an arbitrary approximate joint measurement $\widetilde{M}^{FG}$ of $(F, G)$ and $\epsilon_{1}, \epsilon_{2}\in(0,1]$ satisfying $\epsilon_{1}+\epsilon_{2}\le 1$, there exists a state $\omega\in\Omega$ such that
\[
\begin{aligned}
&D_{W}(\widetilde{M}^{F}, F)\ge \frac{\epsilon_{1}}{2}\ W_{\epsilon_{1}+\epsilon_{2}}(\omega^{F})\\
&D_{W}(\widetilde{M}^{G}, G)\ge \frac{\epsilon_{2}}{2}\ W_{\epsilon_{1}+\epsilon_{2}}(\omega^{G}).
\end{aligned}
\]
\end{cor}
There is also another formulation by means of minimum localization error and $l_{\infty}$ distance.
\begin{thm}
\label{thm_l-distance}
Let $\Omega$ be a transitive state space and its positive cone $V_{+}$ be self-dual with respect to $\langle\cdot,\cdot\rangle_{GL(\Omega)}$, and let $(F, G)$ be a pair of ideal measurements on $\Omega$.
For an arbitrary approximate joint measurement $\widetilde{M}^{FG}$ of $(F, G)$, there exists a state $\omega\in\Omega$ such that
\[
D_{\infty}(\widetilde{M}^{F}, F)+D_{\infty}(\widetilde{M}^{G}, G)\ge LE(\omega^{F})+LE(\omega^{G}).
\]
\end{thm}
\begin{pf}
We can see from \eqref{eq_eigenstate} in Lemma \ref{lem_eigenstates} and the definition of the $l_{\infty}$ distance \eqref{def_l distance} that 
\[
\left|
\left\langle f_{a},\ \frac{f_{a}}{\langle u, f_{a}\rangle}\right\rangle-\left\langle \widetilde{m}^{F}_{a},\ \frac{f_{a}}{\langle u, f_{a}\rangle}\right\rangle
\right|
\le
D_{\infty}(\widetilde{M}^{F}, F)
\]
holds for all $a\in A$, which can be rewritten as
\begin{equation*}
1-\sum_{b\in B}\left\langle \widetilde{m}^{FG}_{ab},\ \frac{f_{a}}{\langle u, f_{a}\rangle}\right\rangle
\le
D_{\infty}(\widetilde{M}^{F}, F),
\end{equation*}
for all $a\in A$. Multiplying both sides by $\langle u, f_{a}\rangle$ and taking the summation over $a$, we have
\[
1-\sum_{a\in A}\sum_{b\in B}\left\langle \widetilde{m}^{FG}_{ab}, f_{a}\right\rangle
\le
D_{\infty}(\widetilde{M}^{F}, F),
\] 
namely
\begin{equation}
\label{ineq_fa2}
1-\sum_{a'\in A}\sum_{b'\in B}\langle u, \widetilde{m}_{a'b'}^{FG}\rangle\left\langle f_{a'},\ \frac{\widetilde{m}_{a'b'}^{FG}}{\langle u, \widetilde{m}_{a'b'}^{FG}\rangle}\right\rangle
\le
D_{\infty}(\widetilde{M}^{F}, F)
\end{equation}
In a similar way, we also have
\begin{equation}
\label{ineq_gb2}
1-\sum_{a'\in A}\sum_{b'\in B}\langle u, \widetilde{m}_{a'b'}^{FG}\rangle\left\langle g_{b'},\ \frac{\widetilde{m}_{a'b'}^{FG}}{\langle u, \widetilde{m}_{a'b'}^{FG}\rangle}\right\rangle
\le
D_{\infty}(\widetilde{M}^{G}, G).
\end{equation}
Since $\sum_{a'\in A}\sum_{b'\in B}\langle u, \widetilde{m}_{a'b'}^{FG}\rangle=1$, \eqref{ineq_fa2} and \eqref{ineq_gb2} give
\begin{equation*}
\begin{aligned}
\sum_{a'\in A}\sum_{b'\in B}\langle u, \widetilde{m}_{a'b'}^{FG}\rangle
\left[
\left(
1-\left\langle f_{a'},\ \frac{\widetilde{m}_{a'b'}^{FG}}{\langle u, \widetilde{m}_{a'b'}^{FG}\rangle}\right\rangle
\right)\right.
+
\left.\left(
1-\left\langle g_{b'},\ \frac{\widetilde{m}_{a'b'}^{FG}}{\langle u, \widetilde{m}_{a'b'}^{FG}\rangle}\right\rangle
\right)
\right]\\
\le D_{\infty}(\widetilde{M}^{F}, F)+D_{\infty}(\widetilde{M}^{G}, G),
\end{aligned}
\end{equation*}
which indicates that there exists a $(a'_{0}, b'_{0})\in A\times B$ satisfying
\begin{align}
\left(
1-\left\langle f_{a'_{0}},\ \frac{\widetilde{m}_{a'_{0}b'_{0}}^{FG}}{\langle u, \widetilde{m}_{a'_{0}b'_{0}}^{FG}\rangle}\right\rangle
\right)
+
\left(
1-\left\langle g_{b'_{0}},\ \frac{\widetilde{m}_{a'_{0}b'_{0}}^{FG}}{\langle u, \widetilde{m}_{a'_{0}b'_{0}}^{FG}\rangle}\right\rangle
\right)\qquad\qquad\notag\\
\le
D_{\infty}(\widetilde{M}^{F}, F)+D_{\infty}(\widetilde{M}^{G}, G)\label{ineq_fa gb2}.
\end{align}
Because 
\[
\omega'_{0}:=\frac{\widetilde{m}_{a'_{0}b'_{0}}^{FG}}{\langle u, \widetilde{m}_{a'_{0}b'_{0}}^{FG}\rangle}
\]
is a state (\eqref{eq_eigenstate0} in Lemma \ref{lem_eigenstates}), we can conclude from \eqref{ineq_fa gb2} and the definition of the minimum localization error \eqref{def_localization error} that
\[
LE({\omega'}_{0}^{F})+LE({\omega'}_{0}^{G})\le D_{\infty}(\widetilde{M}^{F}, F)+D_{\infty}(\widetilde{M}^{G}, G),
\]
which proves the theorem.
\qed\end{pf}

Our theorems above have been proved only for 
a class of theories
 such as finite dimensional classical and quantum theories, and regular polygon theories with odd sides (see Example \ref{eg_CT} - \ref{eg_polygon}). What is essential to the proofs of the theorems is that we can see effects as states (the self-duality), and that every effect of an ideal measurement is an ``eigenstate'' of itself (Lemma \ref{lem_eigenstates}). 
In fact, taking those points into consideration, although it may be a minor generalization, we can demonstrate similar theorems for even-sided regular polygon theories. 
\begin{thm}
\label{thm_for even polygon}
Let $n$ be an even integer. Theorem \ref{thm_error bar}, Corollary \ref{cor_Werner measure}, and Theorem \ref{thm_l-distance} hold also for the $n$-sided regular polygon theory in Example \ref{eg_polygon}.
\end{thm}
\begin{pf}
The proof is done by confirming that the claim of Lemma \ref{lem_eigenstates} holds for even-sided regular polygon theories with modified parametrizations.
We again denote the inner product $\langle \cdot ,\cdot \rangle_{GL(\Omega_{n})}$ by $\langle\cdot ,\cdot \rangle$ in this proof.

In the $n$-sided regular polygon theory with even $n$, if $F=\{f_{a}\}_{a}$ is an ideal measurement, then it is of the form
\begin{equation}
\label{def_even_ideal}
F=\{f_{0}, f_{1}\}
\end{equation}
with
\begin{equation}
\label{def_even_ideal2}
f_{0}=e_{n}^{\mathrm{ext}}(i)\quad\mbox{and}\quad f_{1}=u-e_{n}^{\mathrm{ext}}(i)=e_{n}^{\mathrm{ext}}(i+\frac{n}{2})
\end{equation}
for some $i$ (remember that we do not consider the trivial measurement $F=\{u\}$). Let us introduce an affine bijection
\begin{align}
\label{def_affine bijection}
\psi:=\left(
\begin{array}{ccc}
r_{n} & 0 & 0 \\
0 & r_{n} & 0 \\
0 & 0 & 1
\end{array}
\right)
\end{align}
on $\R^{3}$. Because $(e, \omega)_{E}=(\psi^{-1}(e), \psi(\omega))_{E}$ holds for any $\omega\in\Omega_{n}$ and $e\in\mathcal{E}(\Omega_{n})$, we can consider an equivalent expression of the theory with $\psi\left(\Omega_{n}\right)=:\widehat{\Omega}_{n}$ and $\psi^{-1}\left(\mathcal{E}(\Omega_{n})\right)$ being its state and effect space respectively. The pure states \eqref{def_polygon pure state} and  the extreme effects \eqref{def_polygon pure effect} are modified as
\begin{align}
\omega_{n}^{\mathrm{ext}}(i)&\ \rightarrow\ 
\widehat{\omega}_{n}^{\ \mathrm{ext}}(i):=\psi\left(\omega_{n}^{\mathrm{ext}}(i)\right)
=
\left(
\begin{array}{c}
r_{n}^{2}\cos({\frac{2\pi i}{n}})\\
r_{n}^{2}\sin({\frac{2\pi i}{n}})\\
1
\end{array}
\right)\label{def_polygon pure state2}\\
e_{n}^{\mathrm{ext}}(i)&\ \rightarrow\ 
2\ \widecheck{e}_{n}^{\ \mathrm{ext}}(i):=\psi^{-1}\left(e_{n}^{\mathrm{ext}}(i)\right)
=\frac{1}{2}
\left(
\begin{array}{c}
\cos({\frac{(2i-1)\pi}{n}})\\
\sin({\frac{(2i-1)\pi}{n}})\\
1
\end{array}
\right)\label{def_polygon pure effect2}
\end{align}
respectively, and their conic hull (the positive cone and the internal dual cone) as 
\begin{align*}
V_{+}&\ \rightarrow\ \widehat{V}_{+}:=\psi\left(V_{+}\right)
\\
V^{*int}_{+\langle \cdot ,\cdot \rangle}&\ \rightarrow\ 
\widecheck{V}^{*int}_{+\langle \cdot ,\cdot \rangle}:=\psi^{-1}\left(V^{*int}_{+\langle \cdot ,\cdot \rangle}\right),
\end{align*}
respectively. Note in the equations above that $GL(\Omega_{n})=GL(\widehat{\Omega}_{n})$ and $(\cdot ,\cdot )_{E}=\langle \cdot ,\cdot \rangle_{GL(\Omega_{n})}=\langle \cdot ,\cdot \rangle_{GL(\widehat{\Omega}_{n})}=\langle\cdot ,\cdot \rangle$ hold, and $\omega_{M}=u=\ ^{t}(0,0,1)$ is invariant for $\psi$ (and $\psi^{-1}$). We can also find that a measurement $E=\{e_{a}\}_{a}$ in the original expression is rewritten as $\widecheck{E}:=\{\widecheck{e}_{a}\}_{a}$ with $\widecheck{e}_{a}:=\psi^{-1}(e_{a})$, and that an ideal measurement $F$ in \eqref{def_even_ideal} and \eqref{def_even_ideal2} gives
\begin{equation}
\label{def_even_ideal3}
\widecheck{F}=\{\widecheck{f}_{0}, \widecheck{f}_{1}\}
\end{equation}
with
\begin{equation}
\label{def_even_ideal4}
\widecheck{f}_{0}=\widecheck{e}_{n}^{\ \mathrm{ext}}(i)\quad\mbox{and}\quad \widecheck{f}_{1}=u-\widecheck{e}_{n}^{\ \mathrm{ext}}(i)=\widecheck{e}_{n}^{\ \mathrm{ext}}(i+\frac{n}{2}),
\end{equation}
which is also ideal in the rewritten theory. 
Since 
\begin{equation}
	\label{eq_pf2}
	\left\langle \widecheck{e}_{n}^{\ \mathrm{ext}}(i),\ \frac{\widecheck{e}_{n}^{\ \mathrm{ext}}(i)}{\langle u, \widecheck{e}_{n}^{\ \mathrm{ext}}(i)\rangle}\right\rangle=1
\end{equation}
holds for any $i$ (see \eqref{def_polygon pure effect2}), we can conclude together with \eqref{def_even_ideal3} and \eqref{def_even_ideal4} that any ideal measurements $\widecheck{F}=\{\widecheck{f}_{k}\}_{k=0, 1}$ satisfies
\begin{equation}
\label{eq_pf3}
\left\langle \widecheck{f}_{k},\ \frac{\widecheck{f}_{k}}{\langle u, \widecheck{f}_{k}\rangle}\right\rangle=1.
\end{equation}
On the other hand, it can be seen from \eqref{def_polygon pure state2} and \eqref{def_polygon pure effect2} that $\widehat{V}_{+}$ generated by \eqref{def_polygon pure state2} includes $\widecheck{V}^{*int}_{+\langle\cdot ,\cdot \rangle}$ generated by \eqref{def_polygon pure effect2}, i.e. $\widecheck{V}^{*int}_{+\langle\cdot ,\cdot \rangle}\subset \widehat{V}_{+}$ (see FIG \ref{fig_1}).
\begin{figure}[h]
	\centering
	\includegraphics[scale=0.35]{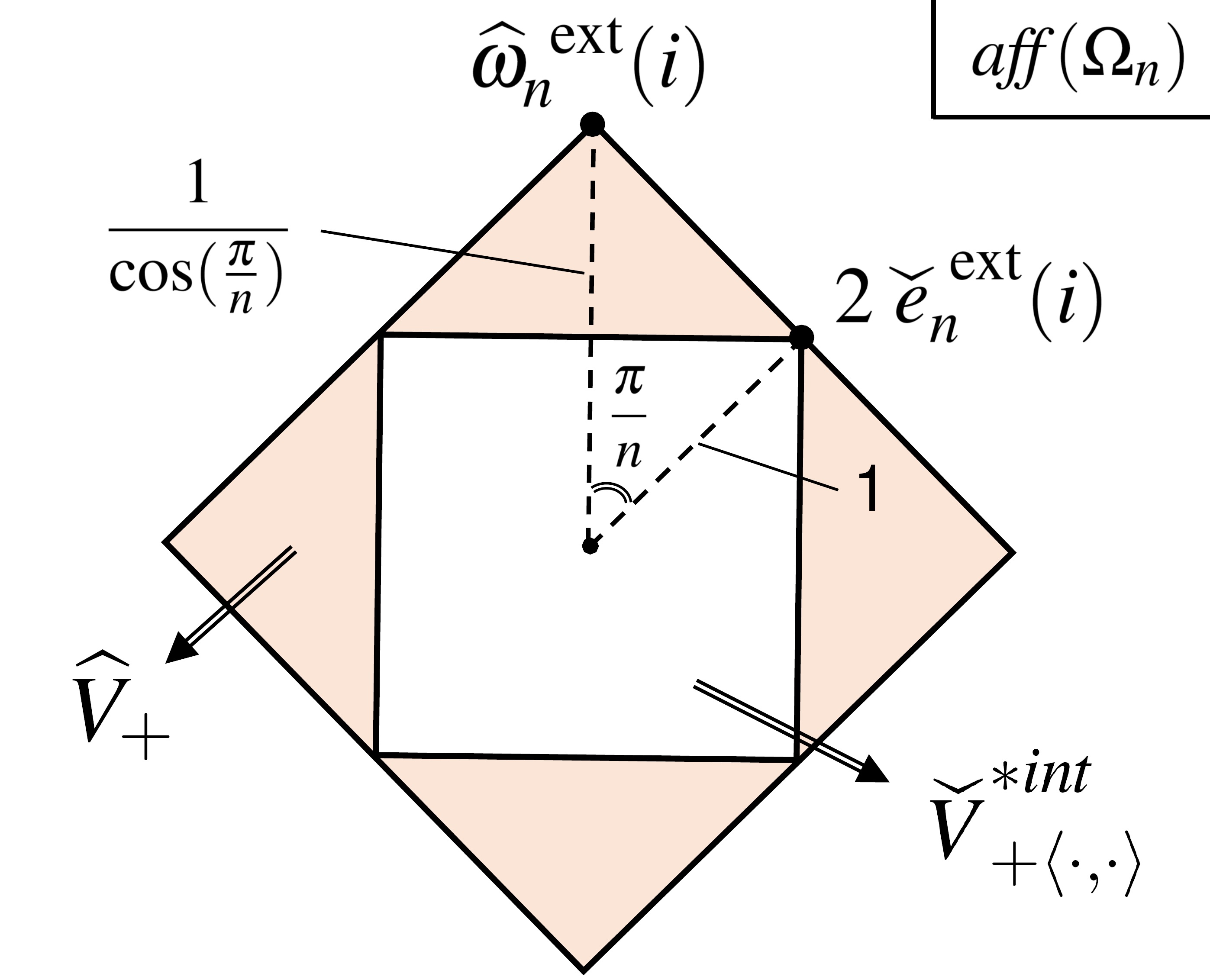}
	\caption{Illustration of $(\mathit{aff}(\Omega_{n})\cap\widehat{V}_{+})=\Omega_{n}$ generated by $\{\widehat{\omega}_{n}^{\ \mathrm{ext}}(i)\}_{i=1}^{n}$ \eqref{def_polygon pure state2}
	and $(\mathit{aff}(\Omega_{n})\cap\widecheck{V}^{*int}_{+\langle\cdot ,\cdot \rangle})$ generated by $\{2\ \widehat{e}_{n}^{\ \mathrm{ext}}(i)\}_{i=1}^{n}$ \eqref{def_polygon pure effect2}
	 for $n=4$.
	 It is observed that
	 $\widecheck{V}^{*int}_{+\langle\cdot ,\cdot \rangle}\subset \widehat{V}_{+}$, which holds also for every even $n$.
	 }
	\label{fig_1}
\end{figure}
Therefore,
\begin{equation}
	\label{eq_pf1}
	\frac{\widecheck{e}}{\langle u,\widecheck{e}\rangle}\in\widehat{\Omega}_{n}
\end{equation}
holds for any effect $\widecheck{e}\in \widecheck{V}^{*int}_{+\langle\cdot ,\cdot \rangle}$.
It follows from \eqref{eq_pf3} and \eqref{eq_pf1} that the claim of Lemma \ref{lem_eigenstates} holds also for even-sided regular polygon theories in a rewritten expression \eqref{def_polygon pure state2} and \eqref{def_polygon pure effect2}.

We also need to confirm that all of our measures \eqref{def_overall width}, \eqref{def_localization error}, \eqref{def_error bar width}, \eqref{def_Werner's measure}, and \eqref{def_l distance} depend only on probabilities, and thus they are invariant for the modification above. For example, for a pair of measurement $M=\{m_{a}\}_{a}$ and $F=\{f_{a}\}_{a}$ on the original state space $\Omega_{n}$, we can see easily from \eqref{def_localization error} and \eqref{def_l distance} that
\begin{align*}
LE(\omega^{F})
&=1-\underset{a\in A}{\max}\ f_{a}(\omega)\\
&=1-\underset{a\in A}{\max}\ \widecheck{f}_{a}(\widehat{\omega})\\
&=LE(\widehat{\omega}^{\widecheck{F}})
\end{align*}
and
\begin{align*}
D_{\infty}(M, F)
&=\underset{\omega\in\Omega_{n}}{\sup}\ \underset{a\in A}{\max}\left|m_{a}(\omega)-f_{a}(\omega)\right|\\
&=\underset{\widehat{\omega}\in\widehat{\Omega}_{n}}{\sup}\ \underset{a\in A}{\max}\left|\widecheck{m}_{a}(\widehat{\omega})-\widecheck{f}_{a}(\widehat{\omega})\right|\\
&=D_{\infty}(\widecheck{M}, \widecheck{F})
\end{align*}
respectively. 
It results in that if Theorem \ref{thm_l-distance} holds in the modified theory, then it holds also in the original theory. 
In fact, by virtue of \eqref{eq_pf3} and \eqref{eq_pf1} (the ``generalized version of Lemma \ref{lem_eigenstates}''), we can repeat the same calculations as in Theorem \ref{thm_l-distance}, and obtain a similar result to it in the modified theory. 
Similar considerations can be adapted also for the other measures, and it proves Theorem \ref{thm_for even polygon}. 
\qed
\end{pf}

\begin{rmk}
	It was claimed in \onlinecite{PhysRevA.101.052104} similarly to our theorems that PURs imply MURs in GPTs. 
	However, the result in \onlinecite{PhysRevA.101.052104} was obtained for a pair of binary (i.e. two-outcome), extreme, sharp, and {\it postprocessing clean} \cite{PhysRevA.97.062102} observables. 
	It is known that any effect of a sharp and postprocessing clean observable is pure and indecomposable, and such observables do not always exist for a GPT \cite{KIMURA2010175,PhysRevA.97.062102}. Actually, the only finite dimensional quantum theory admitting those observables is a qubit system (remember that pure and indecomposable effects correspond to rank-1 projections in finite dimensional quantum theories). On the other hand, although our GPTs are assumed to be transitive and self-dual, or regular polygon theories, our theorems are obtained for more general forms of observables \eqref{def of ideal meas} always possible to be defined.
\end{rmk}

Theorem \ref{thm_l-distance} (and Theorem \ref{thm_for even polygon}) has an application to evaluate the {\it degree of incompatibility} \cite{Busch_2013,PhysRevA.87.052125,PhysRevA.89.022123} of a GPT.

\begin{eg}[Evaluation of degree of incompatibility]
\label{eg_deg of inc}
Suppose that $\Omega$ is an arbitrary state space, and $F$ and $G$ are two-outcome measurements on $\Omega$, namely $F=\{f_{0}, f_{1}\}$ and $G=\{g_{0}, g_{1}\}$, and consider similarly to \eqref{def_appro Z and X} their ``fuzzy'' versions
\begin{equation}
\label{def_appro F and G}
\begin{aligned}
\widetilde{F}^{\lambda}:&=\lambda F+(1-\lambda)\{\frac{u}{2},\ \frac{u}{2}\}=\left\{\lambda f_{0}+\frac{1-\lambda}{2}u,\ \lambda f_{1}+\frac{1-\lambda}{2}u\right\}\\
\widetilde{G}^{\lambda}:&=\lambda G+(1-\lambda)\{\frac{u}{2},\ \frac{u}{2}\}=\left\{\lambda g_{0}+\frac{1-\lambda}{2}u,\ \lambda g_{1}+\frac{1-\lambda}{2}u\right\}
\end{aligned}
\end{equation}
for $\lambda\in[0, 1]$. It is known that we can find a $\lambda_{F, G}\ge\frac{1}{2}$ such that the distorted measurements $\widetilde{F}^{\lambda}$ and $\widetilde{G}^{\lambda}$ in \eqref{def_appro F and G} are jointly measurable for any $\lambda\in[0, \lambda_{F, G}]$, and $\lambda_{\mathrm{opt}}:=\inf_{F, G}\lambda_{F, G}$ can be thought describing the degree of incompatibility of the theory. $\lambda_{\mathrm{opt}}$ has been calculated in various theories: for example, $\lambda_{\mathrm{opt}}=\frac{1}{\sqrt{2}}$ in finite dimensional quantum theories \cite{PhysRevA.87.052125}, and $\lambda_{\mathrm{opt}}=\frac{1}{2}$ in the square theory (a regular polygon theory with $n=4$) \cite{PhysRevA.96.022113}.

To see how Theorem \ref{thm_l-distance} contributes to the degree of incompatibility, we consider the situations in Theorem \ref{thm_l-distance} (and Theorem \ref{thm_for even polygon}) with the marginals $\widetilde{M}^{F}$ and $\widetilde{M}^{G}$ of the approximate joint measurement being $\widetilde{F}^{\lambda}$ and $\widetilde{G}^{\lambda}$ in \eqref{def_appro F and G} for $\lambda\in[0, \lambda_{F, G}]$ respectively. In this case, we can represent the measurement error $D_{\infty}(\widetilde{F}^{\lambda}, F)$ in a more explicit way:
\begin{align}
\label{def_l distance1}
D_{\infty}(\widetilde{F}^{\lambda}, F)
&=\underset{\omega\in\Omega}{\sup}\ \underset{i\in \{0,1\}}{\max}\left|\left(\lambda f_{i}+\frac{1-\lambda}{2}u\right)(\omega)-f_{i}(\omega)\right|\notag\\
&=(1-\lambda)\ \underset{\omega\in\Omega}{\sup}\ \underset{i\in \{0,1\}}{\max}\left|f_{i}(\omega)-\frac{1}{2}\right|\notag\\
&=\frac{1-\lambda}{2},
\end{align}
where we use the relation
\[
\left|f_{0}(\omega)-\frac{1}{2}\right|=\left|(u-f_{1})(\omega)-\frac{1}{2}\right|=\left|f_{1}(\omega)-\frac{1}{2}\right|
\]
and the fact that there is an ``eigenstate'' $\omega_{i}$ for each ideal effect $f_{i}$ satisfying $f_{i}(\omega_{i})=1$ as we have seen in \eqref{eq_eigenstate} or \eqref{eq_pf2}. Therefore, we can conclude from Theorem \ref{thm_l-distance} (and Theorem \ref{thm_for even polygon}) that for any $\lambda\in[0, \lambda_{F, G}]$ and for some state $\omega_{0}$
\begin{align*}
1-\lambda\ge \left(1-\underset{i\in \{0,1\}}{\max}f_{i}(\omega_{0})\right)+\left(1-\underset{j\in \{0,1\}}{\max}g_{j}(\omega_{0})\right)
\end{align*}
holds, that is,
\begin{align}
\label{eq_deg of inc for bi 1}
\lambda_{F, G}\le \underset{\omega\in\Omega}{\max}\left(\underset{i\in \{0,1\}}{\max}f_{i}(\omega)+\underset{j\in \{0,1\}}{\max}g_{j}(\omega)\right)-1
\end{align}
holds, and $\lambda_{\mathrm{opt}}$ can be evaluated by taking the infimum of both sides of \eqref{eq_deg of inc for bi 1} over all two-outcome measurements. 
We remark that the maximum value in the right hand side of \eqref{eq_deg of inc for bi 1} does exist due to the compactness of $\Omega$. 
If we restricted ourselves to regular polygon theories, our inequality \eqref{eq_deg of inc for bi 1} gives unfortunately meaningless bounds $\lambda_{\mathrm{opt}}\le1$ for $n=3, 4, 5, 6$, where we can always find a state $\omega^{\star}$ such that 
\[
\underset{i\in \{0,1\}}{\max}f_{i}(\omega^{\star})+\underset{j\in \{0,1\}}{\max}g_{j}(\omega^{\star})=2
\]
for any pair of ideal measurements $F,G$.
However, when $n\ge7$, we can obtain nontrivial bounds in \eqref{eq_deg of inc for bi 1}.
For example, consider regular polygon theories with $n=4k$ $(k=1, 2, \cdots)$ and let $F$ and $G$ be ideal measurements such that 
the Bloch vectors $\bm{f_{0}}$ and $\bm{g_{0}}$ corresponding to the effects $f_{0}$ of $F$ and $g_{0}$ of $G$ respectively are perpendicular to each other.
These $F$ and $G$ can be regarded as generalizations of the quantum observables $Z=\{\ketbra{0}{0}, \ketbra{1}{1}\}$ and $X=\{\ketbra{+}{+}, \ketbra{-}{-}\}$ respectively because the effects $\ketbra{0}{0}$ and $\ketbra{+}{+}$ (or the vectors $\ket{0}$ and $\ket{+}$) are perpendicular to each other on the Bloch ball.
The right hand side of \eqref{eq_deg of inc for bi 1} for this case was calculated in \onlinecite{Takakura_entropicUR} as
\begin{align}
\label{eq_deg of inc for bi 2}
\underset{\omega\in\Omega}{\max}\left(\underset{i\in \{0,1\}}{\max}f_{i}(\omega)+\underset{j\in \{0,1\}}{\max}g_{j}(\omega)\right)-1
=\left\{
\begin{aligned}
&\frac{r_{n}^{2}}{\sqrt{2}}\quad(n\equiv 4\ (\mbox{mod}\ 8))\\
&\frac{1}{\sqrt{2}}\quad(n\equiv 0\ (\mbox{mod}\ 8))\\
&\frac{1}{\sqrt{2}}\quad(n=\infty).
\end{aligned}
\right.
\end{align}
We can observe that \eqref{eq_deg of inc for bi 2} reproduces properly the quantum bound $\lambda_{Z, X}\le\frac{1}{\sqrt{2}}$.
\end{eg}

\section{Conclusion and Discussion}
\label{sec4}
In our study, although only theories with transitivity and self-duality with respect to a certain inner product were considered, it was revealed that similar quantitative relations between preparation and measurement uncertainty to quantum case \cite{doi:10.1063/1.3614503} hold also in GPTs. Because GPTs considered in this paper include classical, quantum, and other theories such as regular polygon theories, our results can be considered as generalizations of the quantum ones. It is easy to see from the proofs that our theorems can be generalized to the case when three or more measurements are considered. While our assumptions may seem curious, it has been observed in \onlinecite{PhysRevLett.108.130401} that those two conditions are satisfied simultaneously if the state space is {\it bit-symmetric}. There are also researches where they are derived from certain conditions possible to be interpreted physically \cite{Barnum_2014,1367-2630-19-4-043025}. However, considering that our theorems also hold in regular polygon theories with even sides, which are not self-dual, future research is required to investigate whether we can loosen the assumptions.

What is also specific to our main theorems is that their proofs do not require the rules of determining composite systems, while the quantum results of the previous study \cite{doi:10.1063/1.3614503} were proved by means of the maximally entangled state and its ``ricochet property''. It is known that in GPTs there exist ambiguities when constituting the composite system of two systems \cite{PhysRevLett.99.240501,barnum2012teleportation}, but our theorems avoid successfully those difficulties. Future research should reveal the relations between the maximal entanglement and self-duality, which will be a key to generalize our theorems to infinite dimensional cases (remember that the maximally entangled states cannot be defined in infinite dimensional quantum theories such as $\HH=L^{2}(\R)$).

As seen Example \ref{eg_deg of inc}, our theorems also can be considered as yielding via PURs a method for evaluating measurement error, which is in general hard to obtain \cite{10.5555/2017011.2017020}, and it has been discussed that measurement error quantifies the degree of nonlocality \cite{PhysRevA.87.052125,PhysRevA.89.022123}. Although our method turns to be meaningless for theories such as the square theory, where there is no preparation uncertainty, our results will provide an application to understand the nonlocality in GPTs, which is also a future problem.

\begin{acknowledgments}
We would like to thank the anonymous referee for many fruitful comments.
TM acknowledges financial support from JSPS (KAKENHI Grant Number 20K03732).
\end{acknowledgments}

\appendix
\section{Proof of Proposition \ref{prop_ortho repr}}
\label{appA}
\renewcommand{\theequation}{A\arabic{equation}}
\setcounter{equation}{0}
\renewcommand{\thesection}{\Alph{section}}
\setcounter{subsection}{0}

In this part, we give a proof of Proposition \ref{prop_ortho repr}. 
We need the following proposition, which holds without the assumption of the transitivity of $\Omega$.

\begin{propapp}
\label{prop_generalized ortho repr}
For a state space $\Omega$, define a linear map $P_{M}\colon V\rightarrow V$ by 
\[
P_{M}x=\int_{GL(\Omega)} Tx\ d\mu(T).
\]
Then, $P_{M}$ is an orthogonal projection with respect to the inner product $\langle\cdot ,\cdot \rangle_{GL(\Omega)}$, i.e.
\[
P_{M}=P_{M}^{2}\quad\mbox{and}\quad\langle P_{M}x,\ y\rangle_{GL(\Omega)}=\langle x,\ P_{M}y\rangle_{GL(\Omega)}\ \ \mbox{for all}\ x, y\in V.
\]
\end{propapp}
\begin{pf}
We denote the inner product $\langle\cdot ,\cdot \rangle_{GL(\Omega)}$ simply by $\langle\cdot ,\cdot \rangle$ in this proof. 

Let $V_{M}:=\{x\in V\mid Tx=x\ \mbox{for all}\ T\in GL(\Omega)\}$ be the set of all fixed points with respect to $GL(\Omega)$. Then, it is easy to see that $P_{M}x_{M}=x_{M}$ for any $x_{M}\in V_{M}$ and $V_{M}=ImP_{M}$ (in particular $V_{M}$ is a subspace of $V$). Therefore,
\[
P_{M}^{2}x=P_{M}(P_{M}x)=P_{M}x
\]
holds for any $x\in V$, and thus $P_{M}^{2}=P_{M}$. On the other hand, we can observe
\begin{align}
\langle P_{M}x,\ y\rangle
&=\int_{GL(\Omega)}d\mu(T)\  (TP_{M}x,\ Ty)_{E}\notag\\
&=\int_{GL(\Omega)}d\mu(T)\  (P_{M}x,\ Ty)_{E}\notag\\
&=\int_{GL(\Omega)}d\mu(T)\  \left(\int_{GL(\Omega)}d\mu(S)Sx,\ Ty\right)_{E}.\label{pf_app_1}
\end{align}
Let us fix an orthonormal basis $\{w_{i}\}_{i=1}^{N+1}$ of $V$ compatible with the standard Euclidean inner product of $V$, i.e.
\[
(w_{i},\ w_{j})_{E}=\delta_{ij}.
\]
We can consider representing the vector $\int_{GL(\Omega)}d\mu(S)Sx\in V$ by means of the orthonormal basis $\{w_{i}\}_{i}$ as
\[
\int_{GL(\Omega)}d\mu(S)Sx=\sum_{i}\left(w_{i},\ \int_{GL(\Omega)}d\mu(S)Sx\right)_{E}w_{i}.
\]
In fact, the ``$i$th-element'' $\left(w_{i},\ \int_{GL(\Omega)}d\mu(S)Sx\right)_{E}$ is given by (see \onlinecite{KIMURA20141} for more details)
\[
\left(w_{i},\ \int_{GL(\Omega)}d\mu(S)Sx\right)_{E}=\int_{GL(\Omega)}d\mu(S)\ (w_{i},\ Sx)_{E}.
\]
It results in
\begin{align*}
\left(\int_{GL(\Omega)}d\mu(S)Sx,\ Ty\right)_{E}
&=\sum_{i}\left[\int_{GL(\Omega)}d\mu(S) (w_{i},\ Sx)_{E}\right](w_{i},Ty)_{E}\\
&=\int_{GL(\Omega)}d\mu(S) \left[\sum_{i}(Sx,\ w_{i})_{E}(w_{i},Ty)_{E}\right]\\
&=\int_{GL(\Omega)}d\mu(S)(Sx,\ Ty)_{E}.
\end{align*}
Therefore, we obtain
\begin{align*}
&\int_{GL(\Omega)}d\mu(T)\  \left(\int_{GL(\Omega)}d\mu(S)Sx,\ Ty\right)_{E}\\
&\qquad\qquad\qquad\qquad=\int_{GL(\Omega)}d\mu(T)\  \left[\int_{GL(\Omega)}d\mu(S)\ (Sx,\ Ty)_{E}\right]\\
&\qquad\qquad\qquad\qquad=\int_{GL(\Omega)}d\mu(S)\  \left[\int_{GL(\Omega)}d\mu(T)\ (Sx,\ Ty)_{E}\right]\\
&\qquad\qquad\qquad\qquad=\int_{GL(\Omega)}d\mu(S)\  \left(Sx,\ \int_{GL(\Omega)}d\mu(T)Ty\right)_{E},
\end{align*}
where we use Fubini's theorem for the finite Haar measure $\mu$ on $GL(\Omega)$.
\if
Since the vector $\int_{GL(\Omega)}d\mu(S)Sx \in V$ is constructed with its $i$th element
\[
\left(w_{i},\ \int_{GL(\Omega)}d\mu(S)Sx\right)_{E}
\]
in terms of the Euclidean orthonormal basis $\{w_{i}\}_{i=1}^{N+1}$ of $V$ given by 
\[
\int_{GL(\Omega)}d\mu(S)\ (w_{i},\ Sx)_{E},
\]
\fi
We can conclude together with \eqref{pf_app_1} that
\begin{equation*}
\langle P_{M}x,\ y\rangle=\langle x,\ P_{M}y\rangle
\end{equation*}
holds.
\hfill $\Box$
\end{pf}

Proposition \ref{prop_generalized ortho repr} enables us to give an orthogonal decomposition of a vector $x\in V$ such that
\begin{equation}
\label{def_general ortho repr}
x=(\1-P_{M})x+P_{M}x,
\end{equation}
where $(\1-P_{M})x\in V_{M}^{\perp}$ and $P_{M}x\in V_{M}$. When the transitivity of $\Omega$ is assumed, \eqref{def_general ortho repr} is reduced to Proposition \ref{prop_ortho repr}.

\begin{propref}
For a transitive state space $\Omega$, there exists a basis $\{v_{l}\}_{l=1}^{N+1}$ of $V$ orthonormal with respect to the inner product  $\langle\cdot ,\cdot  \rangle_{GL(\Omega)}$ such that $v_{N+1}=\omega_{M}$ and 
\[
x\in\mathit{aff}(\Omega)\iff x=\sum_{l=1}^{N}a_{l}v_{l}+v_{N+1}=\sum_{l=1}^{N}a_{l}v_{l}+\omega_{M}\ (a_{1}, \cdots, a_{N}\in\R).
\]
\end{propref}
\begin{pf}
Since we set $\mathrm{dim}\mathit{aff}(\Omega)=N$, there exists a set of $N$ linear independent vectors $\{v_{l}\}_{l=1}^{N}\subset[\mathit{aff}(\Omega)-\omega_{M}]$ which forms a basis of the $N$-dimensional vector subspace $[\mathit{aff}(\Omega)-\omega_{M}]\subset V$, and we can assume by taking an orthonormalization that they are orthonormal with respect to the inner product $\langle\cdot ,\cdot \rangle$. Hence, $x\in\mathit{aff}(\Omega)$ if and only if it is represented as
\begin{equation}
\label{def_repr of aff}
	x=\sum_{l=1}^{N}a_{l}v_{l}+\omega_{M}\quad(a_{1}, \cdots, a_{N}\in\R).
\end{equation}
Moreover, because of the definition of $\mathit{aff}(\Omega)$, for every $v_{l}\in[\mathit{aff}(\Omega)-\omega_{M}]$ there exist $k\in\mathbb{N}$, real numbers $\{b_{i}\}_{i=1}^{k}$ satisfying $\sum_{i=1}^{k}b_{i}=1$, and states $\{\omega_{i}\}_{i=1}^{k}$ such that $v_{l}=\sum_{i=1}^{k}b_{i}\omega_{i}-\omega_{M}$. By means of  Proposition \ref{def_max mixed state}, we obtain for all $l=1, 2, \cdots, N$
\begin{align}
P_{M}v_{l}
&=\sum_{i=1}^{k}b_{i}P_{M}\omega_{i}-P_{M}\omega_{M}\notag\\
&=\sum_{i=1}^{k}b_{i}\omega_{M}-\omega_{M}=0.\label{app_pf0}
\end{align}
Therefore, because of Proposition \ref{prop_generalized ortho repr}
\begin{align*}
\langle\omega_{M},\ v_{l}\rangle
&=\langle P_{M}\omega_{M},\ v_{l}\rangle\\
&=\langle \omega_{M},\ P_{M}v_{l}\rangle\\
&=0
\end{align*}
holds for all $l=1, 2, \cdots, N$, and we can conclude together with the unit norm of $\omega_{M}$ that $\{v_{1}, \cdots, v_{N}, \omega_{M}\}$ in \eqref{def_repr of aff} forms an orthonormal basis of the $(N+1)$-dimensional vector space $V$ with respect to $\langle\cdot ,\cdot \rangle$ and Proposition \ref{prop_ortho repr} is proved (we can also find that \eqref{def_repr of aff} corresponds to \eqref{def_general ortho repr}).\qed
\end{pf}

\appendix
\setcounter{section}{1}
\section{Proof of Proposition \ref{prop_transitive self-dual}}
\label{appB}
\renewcommand{\theequation}{B\arabic{equation}}
\setcounter{equation}{0}
\renewcommand{\thesection}{\Alph{section}}
\setcounter{subsection}{0}

In this part, we prove Proposition \ref{prop_transitive self-dual}. As we have so far, we let $\Omega$ be a state space, $V_{+}$ be the positive cone generated by $\Omega$, and $GL(\Omega)$ be the set of all state automorphisms on $\Omega$ in the following.
\begin{lemapp}
	\label{lemma0}
	$V_{+\langle\cdot, \cdot\rangle_{GL(\Omega)}}^{*int}$ is a $GL(\Omega)$-invariant set. 
	That is, $T V_{+\langle \cdot, \cdot\rangle_{GL(\Omega)}}^{*int}=V_{+\langle \cdot, \cdot\rangle_{GL(\Omega)}}^{*int}$ for all $T\in GL(\Omega)$.
\end{lemapp}
\begin{pf}
	Let $w\in V_{+\langle\cdot, \cdot\rangle_{GL(\Omega)}}^{*int}$. 
	It holds that $\langle w, v\rangle_{GL(\Omega)}\geq 0$
	for all $v\in V_+$. 
	Because any $T\in GL(\Omega)$ is an orthogonal transformation with respect to $\langle \cdot, \cdot\rangle_{GL(\Omega)}$, we obtain 
	\[
	\langle Tw, v\rangle_{GL(\Omega)}
	= \langle w, T^{-1}v\rangle_{GL(\Omega)} \geq 0
	\]
	for all $v\in V_{+}$. Therefore, $T V_{+\langle \cdot, \cdot\rangle_{GL(\Omega)}}^{*int}\subset V_{+\langle \cdot, \cdot\rangle_{GL(\Omega)}}^{*int}$holds, and a similar argument for $T^{-1}\in GL(\Omega)$ proves the lemma.
\qed
\end{pf}
\begin{lemapp}
\label{lemma1}
Let $(\cdot ,\cdot )$ be an arbitrary inner product on $V$. $V_+$ is self-dual if and only if there exists a linear map $J\colon V\to V$ such that $J$ is strictly positive with respect to $(\cdot ,\cdot )$, i.e. $(x, Jy)=(Jx, y)$ for all $x, y\in V$ and $(x, Jx)>0$ for all $x\in V$, and $J(V_+) = V^{*int}_{+(\cdot ,\cdot )}$. 
\end{lemapp}
\begin{pf}
If part: We introduce an inner product 
$(\cdot, \cdot)_{J}= (\cdot, J\cdot)$.  $V_{+(\cdot,\cdot)_{J}}^{*int}$ 
is written as 
\begin{align*}
V_{+(\cdot,\cdot)_{J}}^{*int}
&=\{v\mid(v, w)_{J} \geq 0, \ ^{\forall} w
\in V_+\}\\
&=\{v\mid(v, Jw)\geq 0, \ ^{\forall} w
\in V_+\}
\\
&= \{v\mid(Jv, w)\geq 0, \ ^{\forall} w\in V_+\}. 
\end{align*}
	Thus, $v\in V_{+(\cdot , \cdot )_{J}}^{*int}$ 
	is equivalent to $Jv\in V_{+(\cdot, \cdot )}^{*int}$. It concludes $V_{+(\cdot,\cdot)_{J}}^{*int}
	=J^{-1}(
	V_{+(\cdot , \cdot )}^{*int}
	)=V_+$. \\
Only if part: Let $V_+$ be self-dual with respect to 
	an inner product $\langle\cdot ,\cdot \rangle$. There exists some $K\colon V\to V$ strictly positive with respect to $(\cdot ,\cdot )$ such that $\langle\cdot ,\cdot\rangle=(\cdot, K\cdot)$. We obtain 
	\begin{align*}
		V_{+}=V_{+\langle\cdot ,\cdot \rangle}^{*int}
		&=\{v|\ \langle v, w\rangle \geq 0, \ ^{\forall} w
		\in V_+\}\\
		&=\{v|\ (v, Kw)\geq 0, \ ^{\forall} w
		\in V_+\}
		\\
		&= \{v|\ (Kv, w)\geq 0, \ ^{\forall} w\in V_+\}
	\end{align*}
	Thus, $v\in V_+=V_{+\langle\cdot ,\cdot \rangle}^{*int}$ is 
	equivalent to $Kv\in V^{*int}_{+(\cdot ,\cdot )}$, i.e.
	 $KV_{+}=V^{*int}_{+(\cdot ,\cdot )}$. Define $J=K$.
	\qed
\end{pf}

In Lemma \ref{lemma1}, we gave a necessary and 
sufficient condition for $V_+$ with an inner product 
$(\cdot ,\cdot )$ to be self-dual. 
The condition was the existence of a strictly positive map 
$J$ satisfying $J(V_+) = 
V_{+(\cdot ,\cdot )}^{*int}$. 
This map $J$ may not be unique. For instance, let us consider 
a classical system in $\R^{2}$ whose extreme points are 
two points $(1,1)$ and $(1,-1)$. The positive cone is a 
``forward lightcone'' $V_+=\{(x_0, x_1)| \ x_0\geq 0, x_0^2 - x_1^2 \geq 0\}$. 
It is easy to see that $V_{+}=V^{*int}_{+(\cdot ,\cdot )_{E}}$ with the standard Euclidean inner product $(\cdot ,\cdot )_{E}$.
However, 
if we choose an orthogonal basis $\{v_{0}, v_{1}\}$ of $\R^{2}$ given by $v_{0}=(1, 1)$ and $v_{1}=(1, -1)$, then
every linear map of the form 
\[
\left(
\begin{aligned}
&v_{0}\\
&v_{1}
\end{aligned}
\right)
\mapsto
\left(
\begin{aligned}
&\lambda_0v_{0}\\
&\lambda_1v_{1}
\end{aligned}
\right)
\]
for $\lambda_0, \lambda_1>0$ (which contains ``Lorentz transformations'' in $1+1$ dimension) is strictly positive and makes $V_{+}$ invariant.
Nevertheless, when $|\Omega^{\mathrm{ext}}|<\infty$, we can demonstrate that such strictly positive maps are ``equivalent'' to each other .

\begin{lemapp}\label{lemma5}
Let $|\Omega^{\mathrm{ext}}|<\infty$.
If a linear map 
	$J:V\to V$ is strictly positive with respect to an inner product $(\cdot ,\cdot )$, i.e. $(x, Jy)=(Jx, y)$ for all $x, y\in V$ and $(x, Jx)>0$ for all $x\in V$, and 
	satisfies $J(V_+)=V_{+}$,   
	then for each $\omega_{}^\mathrm{ext}\in \Omega^\mathrm{ext}$ there exists $\mu(\omega_{}^\mathrm{ext})>0$ such that $J(\omega_{}^\mathrm{ext}) =\mu(\omega_{}^\mathrm{ext}) \omega_{}^\mathrm{ext}$. 
\end{lemapp}
\begin{pf}
	Any $\omega^{\mathrm{ext}}\in\Omega^\mathrm{ext}$ is represented as 
	$\omega^{\mathrm{ext}}=c(\omega^{\mathrm{ext}})w$ with $c(\omega^{\mathrm{ext}}):=\|\omega^{\mathrm{ext}}\|=(\omega^{\mathrm{ext}}, \omega^{\mathrm{ext}})^{1/2}$ and $w$ satisfying $\|w\|=1$. 
	Suppose that there exists a family 
	\[
	\{\omega_k^{\mathrm{ext}}\}_{k=1}^{Z}=\{ c(\omega_k^{\mathrm{ext}})w_{k}\}_{k=1}^{Z}
	\subset \Omega^\mathrm{ext}
	\] 
	such that there is no $\mu(\omega_k^{\mathrm{ext}})>0$ for every $k=1, 2, \cdots, Z$ satisfying 
	$J(\omega_k^{\mathrm{ext}})=\mu(\omega_k^{\mathrm{ext}}) \omega_k^{\mathrm{ext}}$, and define $W:=\{w_{k}\}_{k=1}^{Z}$.
	Since $J$ maps each extreme ray of $V_{+}$ to an extreme ray of $V_{+}$, $J(w_k)$ with $w_{k}\in W$ is proportional to some $\omega^{\mathrm{ext}}\in\Omega^{\mathrm{ext}}$ (remember that an extreme ray of $V_{+}$ is the set of positive scalar multiples of an extreme point of $\Omega$). We can see that $J(w_{k})$ is proportional to some $w_p\in W$ with $p\neq k$ considering that $J(J(w_{k}))=\mu J(w_{k})$ holds if and only if $J(w_{k})=\mu w_{k}$ holds.

	We shall show in the following that there is a $w_{q}\in W$ such that $J(w_{q})\notin W$ despite of the argument above.
	To prove the claim, let us diagonalize $J$. 
	It is written as
	$J=\sum_{n=1}^M \tau_n R_n$, where $\tau_1 
	> \tau_2 >\cdots >\tau_M>0$ and $\{R_n\}_{n=1}^{M}$ are orthogonal projections. 
	We choose $w_1$ so that  
	$0 \neq (w_1, R_1 w_1)
	\geq (w_k, R_1 w_k)$
	for all $w_{k}\in W$. 
	Although such $w_1$ may not be unique, 
	the following argument does not depend on the choice.  
	If it happens that 
	$(w_k, R_1 w_k)=0$
	for all 
	$w_{k}\in W$, 
	we choose $w_1$ so that 
	$0 \neq (w_1, R_2 w_1)
	\geq (w_k, R_2 w_k)$
	for all 
	$w_{k}\in W$. If still $(w_k, R_2 w_k)=0$ for all $w_{k}\in W$, 
	we repeat the argument for $R_3, R_4, \cdots$. 
	For simplicity, we assume hereafter that 
	$(w_1, R_1 w_1)\neq 0$ holds. The general cases can be treated similarly. 
	Let $r_1:=R_1 w_1/\Vert R_1 w_1\Vert\neq0$, then $J$ is written as 
	\begin{align*}
		J= \tau_1\ketbra{r_{1}}{r_{1}}
		+ \tau_1 (R_1 -\ketbra{r_{1}}{r_{1}}) 
		+ \sum_{n\geq 2}\tau_n E_n
		=\tau_1 \hat{R}_0+\tau_1 \hat{R}_1
		+ \sum_{n\geq 2} \tau_n \hat{R}_n,  
	\end{align*} 
	where we define $\hat{R}_0:= \ketbra{r_{1}}{r_{1}}$, $\hat{R}_1:= 
	R_1-\ketbra{r_{1}}{r_{1}}$ and 
	$\hat{R}_n := R_n$ for $n\geq 2$ satisfying $\hat{R}_a\hat{R}_b=\delta_{ab}\hat{R}_a$ for $a,b=0,1,\cdots, M$. 
	Now we consider a vector 
	\begin{align*}
		\frac{J(w_1)}{\Vert J(w_1)\Vert}
		=
		\frac{\tau_1 \hat{R}_{0}w_1
			+ \tau_1 \hat{R}_1 w_1 
			+ \sum_{n\geq 2} \tau_n \hat{R}_n w_1}
		{
			\left(
			\tau_1^2 (w_1, \hat{R}_0w_1)
			+\tau_1^2 (w_1, \hat{R}_1w_1)
			+ \sum_{n \geq 2} \tau_n^2 
			(w_1, \hat{R}_n w_1)
			\right)^{1/2}
		},
	\end{align*}
which must coincide with some $w_p\in W$. Its ``$\hat{R}_{0}$ -element'' can be calculated as
\begin{align}
		&\left( \frac{J(w_1)}{\Vert J(w_1)\Vert}, 
		\hat{R}_0 
		\frac{J(w_1)}{\Vert J(w_1)\Vert}\right)\notag
		\\
		&\qquad\qquad
		=\frac{\tau_1^2 ( w_1, \hat{R}_0 w_1)}
		{
			\tau_1^2 ( w_1, \hat{R}_0 w_1)
			+\tau_1^2 ( w_1, \hat{R}_1w_1)
			+ \sum_{n \geq 2} \tau_n^2 
			( w_1, \hat{R}_n w_1)
		}\notag
		\\&\qquad\qquad
		= \frac{( w_1, \hat{R}_0 w_1)}
		{( w_1, \hat{R}_0 w_1)
			+ ( w_1, \hat{R}_1 w_1 )
			+ 
			\sum_{n=2}^M
			\frac{\tau_n^2}{\tau_1^2}
			( w_1, \hat{R}_n w_1)}.\label{eq_app_R0 element}
	\end{align} 
On the other hand, we can obtain that
\begin{align*}
		&( w_1, \hat{R}_0 w_1)
		+ ( w_1, \hat{R}_1 w_1 )
		+ 
		\sum_{n=2}^M
		\frac{\tau_n^2}{\tau_1^2}
		( w_1, \hat{R}_n w_1)
		\\
		&\qquad\qquad
		<  
		( w_1, \hat{R}_0 w_1)
		+ ( w_1, \hat{R}_1 w_1 )
		+ 
		\sum_{n=2}^M
		( w_1, \hat{R}_n w_1)=1 
	\end{align*}
	because there exists a
	$n\geq 2$ such that $( w_1, \hat{R}_n w_1 )
	\neq 0$ (otherwise $w_{1}=(\hat{R}_{0}+\hat{R}_{1})w_{1}=R_{1}w_{1}$ and thus $J(w_{1})=\tau_{1}w_{1}$ hold, which contradicts $w_{1}\in W$). 
Therefore, \eqref{eq_app_R0 element} results in
	\begin{align*}
	\label{eq_app_R0 element ineq}
			\left( \frac{J(w_1)}{\Vert J(w_1)\Vert}, 
			\hat{R}_0 
			\frac{J(w_1)}{\Vert J(w_1)\Vert}\right)>( w_1, \hat{R}_0 w_1).
	\end{align*}
This observation concludes a contradiction to $J(w_{1})/\|J(w_{1})\|=w_{p}\in W$ because $w_{1}$ satisfies $(w_1, \hat{R}_0 w_1)\geq(w_k, \hat{R}_0 w_k)$ for all $w_{k}\in W$. Overall, we find that 
	every $\omega^{\mathrm{ext}}\in \Omega^{\mathrm{ext}}$ has some
	$\mu(\omega^{\mathrm{ext}})>0$ such that 
	$J(\omega^{\mathrm{ext}})= \mu(\omega^{\mathrm{ext}})\omega^{\mathrm{ext}}$. 
\qed
\end{pf}

\begin{lemapp}\label{prop1}
	Let $|\Omega^{\mathrm{ext}}|<\infty$, and
	suppose that linear maps 
	$J$ and $K$ strictly positive with respect to an inner product 
	$(\cdot ,\cdot )$ satisfy 
	$J(V_+) = 
	K(V_+)=
	V_{+(\cdot ,\cdot )}^{*int}$ (in particular, $V_{+}$ is self-dual). 
	Then, there exists a $\mu(\omega^{\mathrm{ext}})>0$ 
	for each $\omega^{\mathrm{ext}}\in \Omega^{\mathrm{ext}}$ 
	such that $K(\omega^{\mathrm{ext}}) =\mu(\omega^{\mathrm{ext}}) J(\omega^{\mathrm{ext}})$ holds. 
\end{lemapp}
\begin{pf}
As was seen in Lemma \ref{lemma1}, 
the inner products $(\cdot, \cdot)_{J}:= 
(\cdot, J\cdot)$ and $(\cdot, \cdot)_{K}:= 
(\cdot, K\cdot)$ satisfy 
$V_{+(\cdot, \cdot)_J}^{*int} = V_+$ and $V_{+(\cdot, \cdot)_K}^{*int} = V_+$ respectively. 
Because $(\cdot, \cdot)_K$ 
is represented as 
$(\cdot, \cdot)_K=(\cdot, L\cdot)_J$ 
with some linear map $L$ strictly positive with 
respect to $(\cdot, \cdot)_J$, we have for arbitrary $v, w\in V$
\begin{align*}
(v, w)_K=(v, Kw) 
= (v, Lw)_J= (v, JLw),
\end{align*}
and thus $L=J^{-1}\circ K$ holds. On the other hand, $L$ satisfies 
\begin{align*}
	V_{+(\cdot, \cdot)_K}^{*int}
    &= \{v\mid(v, w)_K \geq 0, \ ^{\forall} w\in V_+\}
	\\
	&=\{v\mid(v, Lw)_J \geq 0, \ ^{\forall} w\in V_+\}
	\\
	&=\{v\mid(Lv, w)_J\geq 0, \ ^{\forall} w\in V_+\}
	=L^{-1}(V^{*int}_{+(\cdot, \cdot)_{J}}).
\end{align*}
That is, $L(V_+) =V_+$ holds. Therefore, we can apply Lemma \ref{lemma5} to $L$, and conclude that 
\begin{align*}
L(\omega^{\mathrm{ext}})=\mu(\omega^{\mathrm{ext}})\omega^{\mathrm{ext}}=J^{-1}(K(\omega^{\mathrm{ext}})),
\end{align*}
i.e. $K(\omega^{\mathrm{ext}})=\mu(\omega^{\mathrm{ext}})J(\omega^{\mathrm{ext}})$ holds.
\qed
\end{pf}

\begin{propref1}
Let $\Omega$ be transitive with $|\Omega^{\mathrm{ext}}|<\infty$ and $V_+$ be self-dual with respect to some inner product. There exists a linear bijection $\Xi\colon V\to V$ such that $\Omega':=\Xi\Omega$ is transitive and the generating positive cone $V'_{+}$ is self-dual with respect to $\langle\cdot,\cdot\rangle_{GL(\Omega')}$, i.e.
$V^{'}_+ = V_{+\langle\cdot ,\cdot \rangle_{GL(\Omega')}}^{'*int}$.
\end{propref1}

\begin{pf}
Because of the transitivity of $\Omega$, we can adopt the orthogonal coordinate system of $V$ introduced in Proposition \ref{prop_ortho repr}.
Since $V_+$ is self-dual, there exists a linear map 
	$J\colon V\to V$ strictly positive with respect to $\langle\cdot, \cdot\rangle_{GL(\Omega)}$ such that 
	$J(V_+) = V_{+\langle \cdot, \cdot\rangle_{GL(\Omega)}}^{*int}$ (Lemma \ref{lemma1}). We can assume without loss of generality that $J$ satisfies $\langle\omega_{M}, J\omega_{M}\rangle_{GL(\Omega)}=1$.
	\newpage
	\if
	In fact, for any $v\in V_+$, as $V_+$ is 
	$GL(\Omega)$-invariant $T(v) \in V_+$ follows. 
	Thus $N\circ T(v) \in 
	V_{+, \langle \cdot, \cdot\rangle_{GL(\Omega)}}^{
		*int}$. As $V_{+, \langle \cdot, \cdot\rangle_{GL(\Omega)}}^{
		*int}$ is also $GL(\Omega)$-invariant, $N_T(v)\in 
	V_{+, \langle \cdot, \cdot\rangle_{GL(\Omega)}}^{
		*int}$ holds. Thus we find $N_T(V_+) \subseteq 
	V_{+, \langle \cdot, \cdot\rangle_{GL(\Omega)}}^{
		*int}$. 
	In addition, for any $w \in 
	V_{+, \langle \cdot, \cdot\rangle_{GL(\Omega)}}^{
		*int}$, $v:= T^{-1} \circ N^{-1} 
	\circ T(w) \in V_+$ satisfies 
	$N_T (v)=w$. Thus $N_T(V_+) 
	= V_{+, \langle \cdot, \cdot\rangle_{GL(\Omega)}}^{
		*int}$ is concluded. 
	\fi
Let us introduce 
	\[
	\Omega^{*}:=V^{*int}_{+\langle \cdot, \cdot\rangle_{GL(\Omega)}}\cap[z=1]=\{v\in V^{*int}_{+\langle \cdot, \cdot\rangle_{GL(\Omega)}}\mid\langle v, \omega_{M}\rangle_{GL(\Omega)}=1\},
	\]
	where we identify the ``$\omega_{M}$-coordinate'' with ``$z$-coordinate'' in $V$ and define $[z=1]:=\{x\in V\mid\langle x, \omega_{M}\rangle_{GL(\Omega)}=1\}(=\mathit{aff}(\Omega))$ (see Proposition \ref{prop_ortho repr}). 
	Note that since both $V^{*int}_{+\langle \cdot, \cdot\rangle_{GL(\Omega)}}$ and $[z=1]$
	are $GL(\Omega)$-invariant, $\Omega^{*}$ is 
	also $GL(\Omega)$-invariant. 
    It is easy to demonstrate that $\Omega^{*}$ is convex (and compact), and we denote by $\Omega^{*\mathrm{ext}}$ the set of all extreme points of $\Omega^{*}$. We can also see that $\Omega^{*\mathrm{ext}}$ generates the extreme rays of $V^{*int}_{+\langle \cdot, \cdot\rangle_{GL(\Omega)}}$. Because $J$ satisfying $J(V_{+})=V_{+\langle \cdot, \cdot\rangle_{GL(\Omega)}}^{*int}$ is bijective and maps extreme rays of $V_{+}$ to extreme rays of $V_{+\langle \cdot, \cdot\rangle_{GL(\Omega)}}^{*int}$, it holds that $|\Omega^{*\mathrm{ext}}|=|\Omega^{\mathrm{ext}}|$.
	Thus, there exists 
	a bijection $f\colon\Omega^{\mathrm{ext}} \to \Omega^{*\mathrm{ext}}$ and 
	$\beta(\omega^{\mathrm{ext}})>0$ for each $\omega^{\mathrm{ext}}\in \Omega^{\mathrm{ext}}$
	satisfying $J(\omega^{\mathrm{ext}})=\beta(\omega^{\mathrm{ext}}) f(\omega^{\mathrm{ext}})$.

	For each $T \in GL(\Omega)$, we introduce 
	$J_T:= T^{-1} \circ J \circ T$. It is easy to see that $J_T$ satisfies $J_T(V_+) = V_{+\langle \cdot, 
		\cdot\rangle_{GL(\Omega)}}^
	{*int}$ by virtue of Lemma \ref{lemma0}. Furthermore, $J_T$ is shown to be strictly positive with respect to $\langle \cdot, \cdot\rangle_{GL(
		\Omega)}$ because $T\in GL(\Omega)$ is an orthogonal transformation with respect to $\langle \cdot, \cdot\rangle_{GL(
		\Omega)}$.
	Therefore, applying Lemma \ref{prop1} to $J$ and $J_{T}$, 
	there exists $\mu_T:
	\Omega^{\mathrm{ext}} \to \mathbf{R}_{>0}$ 
	such that $J_T(\omega^{\mathrm{ext}})=\mu_{T}(\omega^{\mathrm{ext}})J(\omega^{\mathrm{ext}})$ for $\omega^{\mathrm{ext}}\in \Omega^{\mathrm{ext}}$, that is,
	\begin{align*}
		J_T(\omega^{\mathrm{ext}})&= \mu_T (\omega^{\mathrm{ext}}) J(\omega^{\mathrm{ext}})\\
		&= \mu_T(\omega^{\mathrm{ext}}) \beta(\omega^{\mathrm{ext}}) f(\omega^{\mathrm{ext}})\\
		&=: \beta_T(\omega^{\mathrm{ext}}) f(\omega^{\mathrm{ext}}),
	\end{align*} 
	where we define $\beta_T(\omega^{\mathrm{ext}}):=\mu_T(\omega^{\mathrm{ext}}) \beta(\omega^{\mathrm{ext}})$.
	We calculate this $\beta_T(\omega^{\mathrm{ext}})$.  
	It holds that 
	\begin{align*}
		J_T(\omega^{\mathrm{ext}}) 
		&=T^{-1}\circ J(T\omega^{\mathrm{ext}})\\
		&= T^{-1} (\beta(T\omega^{\mathrm{ext}}) f(T \omega^{\mathrm{ext}}))\\
		&=\beta(T\omega^{\mathrm{ext}}) T^{-1} f(T\omega^{\mathrm{ext}})\\
		&=\beta_T(\omega^{\mathrm{ext}}) f(\omega^{\mathrm{ext}}).  
	\end{align*}
	This relation shows that $T^{-1} f(T\omega^{\mathrm{ext}})$ is proportional to $f(\omega^{\mathrm{ext}})$. 
	Considering that the $z$-coordinates of $f(T\omega^{\mathrm{ext}})$ and $f(\omega^{\mathrm{ext}})$ are $1$ and that 
	$T^{-1}$ preserves $z$-coordinates, 
	we find that $T^{-1} f(T\omega^{\mathrm{ext}}) =f(\omega^{\mathrm{ext}})$ (equivalently, $f(T\omega^{\mathrm{ext}})= T f(\omega^{\mathrm{ext}})$) holds.
	Consequently, we obtain 
	\begin{align*}
		J_T(\omega^{\mathrm{ext}}) = \beta(T\omega^{\mathrm{ext}}) f(\omega^{\mathrm{ext}}). 
	\end{align*}

	Now we introduce 
	\begin{align*}
		J_{av}:= \frac{1}{|GL(\Omega)|}\sum_{
			T \in GL(\Omega)} J_T. 
	\end{align*}
	We note that $|GL(\Omega)|<\infty$ when $|\Omega^{\mathrm{ext}}|<\infty$ because $|GL(\Omega)|\le|\Omega^{\mathrm{ext}}|\hspace{0.1em}!$.
	$J_{av}$ acts on $\omega^{\mathrm{ext}}\in \Omega^{\mathrm{ext}}$ as 
	\begin{align*}
		J_{av}(\omega^{\mathrm{ext}}) = \frac{1}{|GL(\Omega)|}
		\sum_{T \in GL(\Omega)}\beta(T\omega^{\mathrm{ext}}) \cdot f(\omega^{\mathrm{ext}}) 
		=: C f(\omega^{\mathrm{ext}}),
	\end{align*}
	where $C:=\frac{1}{|GL(\Omega)|}
	\sum_{T \in GL(\Omega)}\beta(T\omega^{\mathrm{ext}})$ is a positive constant which does not depend on the choice of $\omega^{\mathrm{ext}}\in \Omega^{\mathrm{ext}}$ because of the transitivity of $\Omega$.
	Thus, the map satisfies 
	$J_{av}(V_+) = 
	V_{+\langle \cdot, \cdot\rangle_{GL(\Omega)}}^{
		*int}$ 
	since $J_{av}(\Omega^{\mathrm{ext}})
	= C \Omega^{*\mathrm{ext}}$, and is strictly positive with respect to $\langle \cdot, \cdot\rangle_{GL(\Omega)}$ since it is a summation of the strictly positive operators $\{J_{T}\}_{T\in GL(\Omega)}$. 
	Moreover, it satisfies  
	\begin{align*}
		J_{av} \circ T = T \circ J_{av}. 
	\end{align*}
	for any $T \in GL(\Omega)$.
	We thus find that $J_{av} \circ P_{M} = P_{M} \circ J_{av}$ holds for the orthogonal projection $P_{M}$ introduced in Proposition \ref{prop_generalized ortho repr}. In fact, 
	\begin{align*}
	J_{av}(P_{M}x)
	&=\frac{1}{|GL(\Omega)|}J_{av}\left(
	\sum_{T \in GL(\Omega)}Tx\right)\\
	&=\frac{1}{|GL(\Omega)|}
	\sum_{T \in GL(\Omega)}T(J_{av}x)\\
	&=P_{M}(J_{av}x)
	\end{align*}
	holds for all $x\in V$.
	Therefore, $J_{av}$ is decomposed into two parts as 
	\begin{align}
	\label{eq_ortho_decomp}
	J_{av}= P_{M}\circ J_{av} \circ P_{M} + 
	P_{M}^{\perp} \circ J_{av} \circ P_{M}^{\perp},
	\end{align}
	where $P_{M}^{\perp}=\1-P_{M}$.
	We note that $V_{M}^{\perp}=ImP_{M}^{\perp}=[\mathit{aff}(\Omega)-\omega_{M}]=\R^{N}$ and $\mathrm{dim}\ V_{M}=\mathrm{dim}\ ImP_{M}=1$ hold by virtue of Proposition \ref{prop_ortho repr}.
	Therefore, the first part of \eqref{eq_ortho_decomp} is proportional to $\1_{V_{M}}=\1_{z}=P_{M}$, and because we set $\langle\omega_{M}, J\omega_{M}\rangle_{GL(\Omega)}=1$ and thus
\begin{align*}
\left\langle \omega_{M},\ P_{M}\circ J_{av} \circ P_{M}\omega_{M}\right\rangle_{GL(\Omega)}
&=\langle \omega_{M}, J_{av} \omega_{M}\rangle_{GL(\Omega)}\\
&=\langle \omega_{M}, P_{M}J \omega_{M}\rangle_{GL(\Omega)}\\
&=\langle \omega_{M}, J \omega_{M}\rangle_{GL(\Omega)}\\
&=1\\
&=\langle \omega_{M}, P_{M}\omega_{M}\rangle_{GL(\Omega)}
\end{align*}
	holds, it is proved that 
	\[
	P_{M}\circ J_{av} \circ P_{M}=P_{M}.
	\]
	Let us examine the second part. 
	Suppose that there exists a nonzero $x\in V_{M}^{\perp}$ such that $Tx=x$ for all $T\in GL(\Omega)$. Then, $P_{M}x=x\neq0$ holds, and it contradicts to \eqref{app_pf0}. Thus, we can find that $GL(\Omega)$ acts irreducibly on $V_{M}^{\perp}$, that is, only $\{0\}$ and $V_{M}^{\perp}=\R^N$ itself are 
	invariant subspaces. It concludes that $P_{M}^{\perp} J_{av} P_{M}^{\perp}$,
	which commutes with every element in $GL(\Omega)$,
	is proportional to $\1_{V_{M}^{\perp}}=\1_{\R^{N}}=P_{M}^{\perp}$ due to Schur's lemma. 
	Consequently, we obtain for some $\xi >0$
	\begin{align*}
		J_{av} = P_{M} + \xi P_{M}^{\perp},
	\end{align*}
	and thus
	\begin{align}
	\label{app_pf1}
	J_{av}(V_{+})
	&=(P_{M} + \xi P_{M}^{\perp})(V_{+})
	=V_{+\langle \cdot, \cdot\rangle_{GL(\Omega)}}^{*int}.
	\end{align}	
	
	Let us introduce a linear bijection 
	\[
	\Xi:=\sqrt{J_{av}}=P_{M}+\sqrt{\xi}P_{M}^{\perp},
	\] 
	strictly positive with respect to $\langle \cdot, \cdot\rangle_{GL(\Omega)}$, and define $\Omega':=\Xi\Omega$. It is easy to check that the positive cone $V_{+}'$ generated by $\Omega'$ is given by $V_{+}'=\Xi V_{+}$, and $GL(\Omega')=\Xi GL(\Omega)\Xi^{-1}=GL(\Omega)$ (moreover, the unique maximally mixed state of $\Omega'$ is still $\omega_{M}$). In addition, we can find that
	\begin{align*}
	V_{+\langle \cdot, \cdot\rangle_{GL(\Omega')}}^{'*int}
	&=\{v\mid \langle v, w'\rangle_{GL(\Omega)}\ge 0,\ ^{\forall}w'\in V'_{+}
	\}\\
	&=\{v\mid \langle v, \Xi w\rangle_{GL(\Omega)}\ge 0,\ ^{\forall}w\in V_{+}
	\}\\
	&=\Xi^{-1}V_{+\langle \cdot, \cdot\rangle_{GL(\Omega)}}^{*int}.
	\end{align*}
	holds.
	Since \eqref{app_pf1} can be rewritten as
	\[
	\Xi V_{+}=\Xi^{-1}V_{+\langle\cdot ,\cdot \rangle_{GL(\Omega)}}^{*int},
	\]
we can conclude
\[
V^{'}_+ = V_{+\langle\cdot ,\cdot \rangle_{GL(\Omega')}}^{'*int}.
\]
\qed
\end{pf}

	\if
	We employ R-representation. We may rescale 
	the coordinate so that every
	pair $v,w \in \Omega$ satisfies
	$\langle v, w\rangle_{GL(\Omega)}>0$. 
	Let us introduce 
	$\Omega^{*}:= V^{*int}_{+, \langle \cdot, \cdot\rangle_{GL(\Omega)}}\cap [z=1]$. 
	As $V^{*int}_{+, \langle \cdot, \cdot\rangle_{GL(\Omega)}}$ 
	is $GL(\Omega)$-invariant, $\Omega^{*}$ is 
	also $GL(\Omega)$-invariant. 
	Thus $\Omega^{*ext}=\{Tw_0| T\in GL(\Omega)\}$ holds, 
	where $w_0$ is a point on a hyperplane $[z=1]$. 
	Moreover, because $N$ is bijective, 
	$|\Omega^{*ext}|=|\Omega^{ext}|$ follows.  
	Each point in $\Omega^{ext}$ is mapped by $N$ 
	to a point which is proportional to 
	a point in $\Omega^{*ext}$. 
	As $N$ is strictly positive it can be diagonalized as 
	$N= \sum_{n} \lambda_n E_n$, where $\lambda_n >0$ 
	and $E_n$ is a projection. 
	Each $v_j \in \Omega^{ext}$ is mapped as 
	\begin{eqnarray*}
		Nv_j = \sum_n \lambda_n E_n v_j,  
	\end{eqnarray*}
	which must be proportional to some $w_j\in \Omega^{*ext}$. 
	That is, there exists $\mu_j>0$ such that 
	\begin{eqnarray*}
		\sum_n \lambda_n E_n v_j = \mu_j w_j 
		= \sum_n \mu_j E_n w_j. 
	\end{eqnarray*}
	Operating $E_m$ to the above equation, 
	we obtain 
	\begin{eqnarray*}
		\lambda_m E_m v_j = \mu_j E_m w_j.  
	\end{eqnarray*}
	It implies that for $m$ with $E_m w_j =0$, 
	$E_m v_j=0$ holds as $\lambda_m, \mu>0$. 
	It implies that 
	for $m$ with $E_m w_j \neq 0$, 
	both $
	\lambda_m = \mu_j >0$ and 
	$E_m v_j = E_m w_j$ holds. 
	For $m$ with $E_m w_j =0$, 
	either $\lambda_m =0$ or $E_m v_j =0$ must hold. 
	Thus we obtain 
	\begin{eqnarray*}
		v_j = w_j + \sum_{m: E_m w_j =0} 
		E_m v_j.
	\end{eqnarray*}
	Thus we can conclude that 
	$v_j = w_j$ and $\lambda_m =\mu$ for all $m$. 
	\fi

\begin{rmk**}
In the case of $|\Omega^{\mathrm{ext}}|=\infty$, 
there exists a counterexample of Lemma \ref{lemma5}. 
\if
Let us consider $\Omega=\ ^{t}\{(\bm{x}, 1)\in \R^4\mid |\bm{x}| \leq 1\}$ (Bloch ball),  
whose extreme point set is 
$\Omega^{\mathrm{ext}}=\{\bm{x}\mid
\bm{x}\in \mathbf{R}^3, |\mathbf{x}|=1\}$. 
This $\Omega^{ext}$ is transitive 
with respect to $GL(\Omega) =O(3)$.  
Let us embed $\Omega$ into $\mathbf{R}^4$ 
so that $\Omega \subset [x_0=1]$. 
\fi
Let us consider a state space
\[
\Omega=\{\ ^t(
1,\bm{x})=\ ^t(
1,x_1,x_2,x_3)\in\R^{4}\mid |\bm{x}|^{2}=x_1^2 + x_2^2 + x_3^2 \leq 1\}
\]
 (the Bloch ball).
$\Omega$ defines a corresponding positive cone $V_+$ 
as
\[
V_+=\{x\in\R^{4}\mid x_0^2 - |\bm{x}|^2 \geq 0, 
x_0\geq 0\},
\]
which can be identified with a forward light cone
of a Minkowski spacetime.
We examine a pure Lorentz transformation 
$\Lambda$ defined for $\lambda \in \R$
as
\begin{eqnarray*}
	\Lambda= \left[
	\begin{array}{cccc}
		\cosh \lambda& \sinh \lambda& 0& 0\\
		\sinh \lambda& \cosh \lambda& 0&0\\
		0&0&1&0\\
		0&0&0&1
	\end{array}
	\right].
\end{eqnarray*}
It is easy to prove that this $\Lambda$ is strictly positive.
\if
as it holds that for any $x
\in \mathbf{R}^4$ 
\begin{eqnarray*}
	(x, \Lambda x) = \frac{1}{2}
	(e^{\lambda} (x_0+x_1)^2 
	+2^{-\lambda} (x_0-x_1)^2) 
	+x_2^2 + x_3^2 \geq 0. 
\end{eqnarray*}
\fi
Since the pure Lorentz transformation 
preserves the Minkowski metric, 
it satisfies $\Lambda(V_+)=V_+$. 
However, $\Lambda$ transforms an extreme point 
$x=\ ^{t}(1,0,1,0)$ to 
\[
\Lambda(x)=\ ^{t}(\cosh \lambda, 
\sinh \lambda, 1,0),
\]
which is not proportional to 
$x$. Investigating whether Proposition \ref{prop_transitive self-dual} still holds when $|\Omega^{\mathrm{ext}}|=\infty$ is a future problem. 
\end{rmk**}

\section{Proof of Proposition \ref{prop_error bar and Werner}}
\label{appC}
\renewcommand{\theequation}{C\arabic{equation}}
\setcounter{equation}{0}
\renewcommand{\thesection}{\Alph{section}}
\setcounter{subsection}{0}

In this appendix, we prove Proposition \ref{prop_error bar and Werner} given in subsection \ref{sec3-2}.
\begin{propref2}
Let $(A,d_{A})$ be a finite metric space, and $F=\{f_{a}\}_{a\in A}$ and $\widetilde{F}=\{\widetilde{f}_{a}\}_{a\in A}$ be an ideal and general measurement respectively. 
Then,
\[
\mathcal{W}_{\epsilon}(\widetilde{F}, F)\le\frac{2}{\epsilon}D_{W}(\tilde{F}, F)
\]
holds for $\epsilon\in(0, 1].$
\end{propref2}
\begin{pf}
Let us define $n:=\frac{D_{W}(\tilde{F}, F)}{\epsilon}$ for $\epsilon\in(0, 1]$, and consider for $a\in A$ a state $\omega\in\Omega$ satisfying $f_{a}(\omega)=1$. 
Remember that such state does exist for every $a\in A$ because $F$ is ideal.
We also define a function $h_{n}$ on $A$ as
\begin{align*}
h_{n}(x):=
\left\{
\begin{aligned}
&n-d_{A}(x, a) &&(d(x, a)\le n)\\
&0 &&(d(x, a)>n).
\end{aligned}
\right.
\end{align*}
It can be seen that
\[
|h_{n}(x_{1})-h_{n}(x_{2})|\le d_{A}(x_{1}, x_{2})
\]
holds for $x_{1}, x_{2}\in A$, and thus we can obtain from the definition of $D_{W}(\tilde{F}, F)$ \eqref{def_Werner's measure}
\[
\left|(\tilde{F}[h_{n}])(\omega)-(F[h_{n}])(\omega)\right|\le D_{W}(\tilde{F}, F).
\]
It results in 
\begin{equation}
\label{eq_appC}
\left|(\tilde{F}[g_{n}])(\omega)-(F[g_{n}])(\omega)\right|\le \frac{D_{W}(\tilde{F}, F)}{n}=\epsilon,
\end{equation}
where we set $g_{n}:=h_{n}/n$.
Since it holds that $g_{n}(x)\le\chi_{O_{d_{A}}(a;\ 2n)}(x)\le 1$ for all $x\in A$, where $\chi_{O_{d_{A}}(a;\ 2n)}$ is the indicator function of the ball $O_{d_{A}}(a;\ 2n)=\{x\in A\mid d_{A}(x, a)\le n\}$, and 
\[
(F[g_{n}])(\omega)=\sum_{x\in A} g_{n}(x)f_{x}(\omega)=g_{n}(a)f_{a}(\omega)=1
\]
because $f_{a}(\omega)=1$, \eqref{eq_appC} can be rewritten as
\[
1-(\tilde{F}[\chi_{O_{d_{A}}(a;\ 2n)}])(\omega)\le \epsilon,
\]
that is,
\begin{equation}
\label{eq_appC1}
\sum_{x\in O_{d_{A}}(a;\ 2n)} \widetilde{f}_{x}(\omega)\ge 1-\epsilon.
\end{equation}
\eqref{eq_appC1} holds for all $a\in A$ and all $\omega\in \Omega$ such that $f_{a}(\omega)=1$, and thus 
\[
2n=\frac{2}{\epsilon}D_{W}(\tilde{F}, F)\ge\mathcal{W}_{\epsilon}(\widetilde{F}, F)
\] 
is concluded (see the definition of $\mathcal{W}_{\epsilon}(\widetilde{F}, F)$ \eqref{def_error bar width}).
\qed
\end{pf}


\bibliographystyle{aipnum4-1} 
\bibliography{ref_URsinGPT}

\end{document}